\documentclass[journal]{IEEEtran}
\usepackage{amsmath}
\usepackage{amssymb}
\usepackage{graphicx}
\usepackage{subfig}
\usepackage{float}
\usepackage{cite}
\usepackage{diagbox}

\usepackage{color}
\usepackage{amsthm}
\newtheorem{theorem}{Theorem}
\newtheorem{remark}{Remark}

\newtheorem{lemma}{Lemma}

\usepackage[ruled,vlined]{algorithm2e}
\usepackage{multirow}
\ifCLASSINFOpdf
\else
\fi

\begin{document}
\title{Privacy-Aware Time-Series Data Sharing \\ with Deep Reinforcement Learning}

\author{Ecenaz Erdemir,~\IEEEmembership{Student Member,~IEEE,}
        Pier Luigi Dragotti,~\IEEEmembership{Fellow,~IEEE,}
        \\ and~Deniz G\"{u}nd\"{u}z,~\IEEEmembership{Senior Member,~IEEE}
\thanks{This work was partially supported by the European Research Council (ERC) through project BEACON (No. 677854).}

\thanks{The authors are with the Department
of Electrical and Electronic Engineering, Imperial College London, London SW7 2AZ, U.K., (e-mail: e.erdemir17@imperial.ac.uk; p.dragotti@imperial.ac.uk; d.gunduz@imperial.ac.uk).}
}


\maketitle

\begin{abstract}
Internet of things (IoT) devices are becoming increasingly popular thanks to many new services and applications they offer. However, in addition to their many benefits, they raise privacy concerns since they share fine-grained time-series user data with untrusted third parties.
In this work, we study the privacy-utility trade-off (PUT) in time-series data sharing. Existing approaches to PUT mainly focus on a single data point; however, temporal correlations in time-series data introduce new challenges. Methods that preserve the privacy for the current time may leak significant amount of information at the trace level as the adversary can exploit temporal correlations in a trace.
We consider sharing the distorted version of a user's true data sequence with an untrusted third party. We measure the privacy leakage by the mutual information between the user's true data sequence and shared version. We consider both the instantaneous and average distortion between the two sequences, under a given distortion measure, as the utility loss metric. To tackle the history-dependent mutual information minimization, we reformulate the problem as a Markov decision process (MDP), and solve it using asynchronous actor-critic deep reinforcement learning (RL). We evaluate the performance of the proposed solution in location trace privacy on both synthetic and GeoLife GPS trajectory datasets. For the latter, we show the validity of our solution by testing the privacy of the released location trajectory against an adversary network.
\end{abstract}

\IEEEpeerreviewmaketitle

\vspace{-0.6cm}
\section{Introduction}
\IEEEPARstart{R}{ecent} advances in Internet of things (IoT) devices have increased the variety of services they provide, such as health monitoring, financial analysis, weather analysis, location-based services (LBSs) and smart metering. Moreover, the integration of some IoT devices with social networks has encouraged the users to share their personal data to obtain useful information from these social platforms. While the users can receive hotel, restaurant and product recommendations from Facebook, Twitter or YouTube when they share their location information, they can also benefit from the personalized dietary tips as a result of sharing their Fitbit activity. However, fine-grained time-series data collected by IoT devices contain sensitive confidential information about the user. Account balance, biomedical measurements, location trace, weather forecast and smart meter readings are typical examples of time-series data which carry sensitive personal information. For instance, a malicious third party can derive an individual's frequently visited destinations, financial situation or social relationships using the shared location information\cite{LPsurvey}. Using non-intrusive load monitoring techniques on smart meter data, an eavesdropper can deduce the user's presence at home, disabilities and even political views due to the TV channel the user is watching \cite{GiulioSPmag}. Besides all, the most sensitive private information, such as patient history, chronic diseases and psychological state, can be revealed by health monitoring systems \cite{ECG,Health}. Therefore, time-series data privacy has been an important concern, and there is an increasing pressure from consumers to keep their data traces private against malicious attackers or untrusted service providers (SPs), while preserving the utility obtained from these IoT services. Our goal in this paper is to study the fundamental privacy-utility trade-off (PUT) when sharing sensitive time-series data.

\vspace{-0.2cm}
\subsection{Related Work}
Time-series data privacy and its applications to various domains have been extensively studied \cite{Timeseries,Health_kanon1,Health_kanon2,Health_DifP,Crypto,InfoTheo_annon,InfoTheo_anon_obfus,Shokri_kanony,Shokri_single,InfoTheo_single,SM_DifP1,SM_DifP2,GiulioRED_ESD,Ravi,Trace1,Trace2,Trace3,WIFS,Giulio,ICASSP,BizBook}. A large body of research has focused on protecting the privacy of a single data point, e.g., the current sensitive measurement \cite{Shokri_kanony,Shokri_single,InfoTheo_single,SM_DifP1,SM_DifP2,GiulioRED_ESD}. However, the temporal relations in time-series data requires going beyond single data point privacy. Individual measurements taken at each time instance, such as electrocardiogram (ECG), body temperature, location, account balance and smart meter readings, are highly correlated and the strategies focusing on the current data privacy might reveal sensitive information about the past or future measurements. 

Differential privacy (DP), k-anonymity and information theoretic metrics are commonly used as privacy measures \cite{Timeseries,Health_kanon1,Health_kanon2,Health_DifP,Crypto,Shokri_kanony,Shokri_single,InfoTheo_single,InfoTheo_annon,InfoTheo_anon_obfus,SM_DifP1,SM_DifP2,GiulioRED_ESD,Ravi,Trace1,Trace2,Trace3,WIFS,ICASSP,Giulio,BizBook}.
By definition, DP prevents the SP from inferring the current sensitive data of the user, even if the SP has the knowledge of all the remaining private data points. K-anonymity ensures that a sensitive data is indistinguishable from at least $k-1$ other data points. However, DP and k-anonymity are meant to ensure the privacy of a single data point in time. Group-DP tackles this issue by applying DP for each user; however, keeping a large number of points private causes high utility loss. In \cite{ShokriQuantify}, it is stated that these are not appropriate measures for location trace privacy since temporal correlations are not taken into account.

As an intermediate framework between DP which assumes complete independence, and group-DP which assumes complete correlation, \textit{pufferfish privacy} considers low average temporal correlations in time-series data \cite{puff1,puff2,puff4}.
In \cite{puff1}, the location release mechanism assumes a hidden Markov model for actual locations. A Bayesian belief on the true current location is updated at each time by observing the noisy location, which is generated via a differentially private method, e.g., by adding independent and identically distributed (i.i.d) random noise drawn from Laplace distribution. However, this work does not focus on protecting the privacy of a trace or trajectory as the authors mention in Section 3.2 in \cite{puff1}. Instead, the mechanism releases random locations around the possible current location, which might preserve the privacy of the current location while revealing the information about future locations. The reason is that the privacy loss is not measured between the true and released traces, but in the neighborhood of each individual true location. A generalization is proposed in \cite{puff2}, which introduces a Markov blanket mechanism assuming that the temporal correlations decrease as the distance between two nodes increases. To hide the effect of a node on the result of a query under the DP framework, the mechanism adds noise which is determined by the number of nearby nodes. In \cite{puff4}, continuous aggregate location release is considered in a pufferfish privacy framework under temporal correlations modeled as a Markov chain. This approach takes into account a certain number of steps forward and backward, while minimizing the differential privacy loss of the current location. Hence, the accumulating privacy loss of DP mechanism is limited to a level determined by the number of forward and backward steps.
However, \cite{puff2} and \cite{puff4} do not take into account trajectory privacy for reasons similar to those in \cite{puff1}.

Several other papers on DP and k-anonymity consider temporal correlations. In \cite{Health_DifP}, physiological measurements are obfuscated before reporting to an SP for PUT. Instead of the entire time-series history, a selected temporal section of the sensor data is considered, and solved by using dynamic program and greedy algorithm. The work in \cite{InfoTheo_annon} focuses on keeping the user identity private in a location privacy setting by performing random permutation on a set of multiple users. However, the users might still be re-identified when attackers have access to auxiliary information. In \cite{InfoTheo_anon_obfus}, authors improve this approach by considering both user identity and location privacy and merging anonymization with obfuscation. However, the risk of re-identification of the user by the adversary still exists and privacy gain by obfuscation depends highly on the number of users. In \cite{SM_DifP2}, DP in a smart meter with a rechargeable battery is achieved by adding noise to the meter readings before reporting to an SP. In order to guarantee DP, the perturbation must be independent of the battery state of charge. However, for a finite capacity battery, the energy management system cannot provide the amount of noise required for preserving privacy.

On the other hand, information-theoretic privacy considers the statistics of the entire time-series in terms of temporal correlations, and study privacy mechanisms that allow arbitrary stochastic transformations of data samples, rather than being limited to addition of noise of a specific form. This is the biggest advantage of information-theoretic privacy over pufferfish privacy where only some degree of temporal correlations are taken into account, and a fixed type of i.i.d. random noise is added for privacy. In \cite{Ravi}, the authors introduce location distortion mechanisms to keep the user's trajectory private, measuring the privacy by the mutual information between the true and released traces, under a constraint on the average distortion between the two. The true trajectory is assumed to form a Markov chain. Due to the computational complexity of history-dependent mutual information optimization, authors propose bounds which take only the current and one-step past locations into account. However, due to temporal correlations in the trajectory, the optimal distortion introduced at each time instance depends on the entire distortion and location history. Hence, the proposed bounds do not guarantee optimality.

In \cite{AshishInfoTheoP}, a smart metering system is considered assuming Markovian energy demands. Privacy is achieved by filtering the energy demand with the help of a rechargeable battery. Information theoretic privacy problem is formulated as a Markov decision process (MDP), and the minimum leakage is obtained numerically through dynamic programming, while a single-letter expression is obtained for an i.i.d. demand. This approach is extended to the scenario with a renewable energy source in \cite{Giulio}. In \cite{ParvMDP}, privacy-cost trade-off is examined with an RB. Due to Markovian demand and price processes, the problem is formulated as a partially observable MDP with belief-dependent rewards ($\rho$-POMDP), and solved by dynamic programming for infinite-horizon. In \cite{ICASSP}, the PUT is characterized numerically by dynamic programming for a special energy generation process. 

In \cite{Erdogdu_TimeSeriesISIT}, PUT of time-series data is considered in both online and offline setting. In the scenario, a user continuously releases data samples which are correlated with its private information, and in return obtains utility from a SP. The proposed schemes are cast as convex optimization problems and solved under hidden Markov model assumption. The simulation results are provided for binary time-series data for a finite time horizon. However, the dimensions of the optimization problems in both schemes grow exponentially with time and the number of sample states. Therefore, in a setting when fine-grained sensor data is considered for a long time horizon, computational complexity of the proposed schemes is very high.

\vspace{-0.3cm}
\subsection{Contributions}
In this work, we consider the scenario in which the user measures time-series data (e.g., location, heartbeat, temperature or energy consumption) generated by a first-order Markov process through an IoT device, and periodically reports a distorted version of her true data to an untrusted SP to gain utility. We assume that the true data becomes available to the user in an online manner. We use the mutual information between the true and distorted data sequences as a measure of privacy loss, and measure the utility of the reported data by a specific distortion metric between the true and distorted samples. For the PUT, we introduce an online private data release policy (PDRP) that minimizes the mutual information while keeping the distortion below a certain threshold. We consider both instantaneous and average distortion constraints. We consider data release policies which take the entire released data history into account, and show its information theoretic optimality.
To tackle the complexity, we exploit the Markovity of the user's true data sequence, and recast the problem as a Markov decision process (MDP). After identifying the structure of the optimal policy, we use advantage actor-critic (A2C) deep reinforcement learning (RL) framework as a tool to evaluate our continuous state and action space MDP numerically. To the best of our knowledge, this is the first time deep RL tools are used to optimize information theoretic time-series data privacy.

The performances of the proposed PDRPs are examined in two specific scenarios: In the first scenario, synthetic location traces are generated considering a user moving in a grid-world with a known Markov mobility pattern. In the second scenario, we use GPS traces of a user from GeoLife dataset \cite{GeoLife,GeoLife2}. For the average distortion constrained case, the proposed PDRP is compared with a myopic location data release mechanism \cite{Ravi}. While the privacy leakage of the considered PDRPs can be evaluated for the synthetic dataset, this cannot be done for the GeoLife trace since we do not know the true statistics of this dataset. Instead, we compare the privacy achieved by
the proposed and myopic policies using an adversary which predicts the current location of the user from the past released locations. The adversary is represented by a long short-term memory (LSTM) predictor. The performances of the proposed policies are tested under various adversary memory sizes.

This paper extends the theoretical approach in our previous work on PUT for location sharing \cite{WIFS}.
Our contributions are summarized as follows:
\begin{itemize}
    \item We propose a simplified PDRP by exploiting the Markov property of the user's true data sequences. Then, we prove the information theoretic optimality of the simplified strategy.
    \item We recast the information theoretic time-series data PUT problem as an MDP and evaluate the optimal PDRP numerically using advantage actor-critic deep RL.
    \item We apply the obtained information-theoretically optimal PDRP on the location trace privacy problem, and evaluate its performance under instantaneous and average distortion constraints using both synthetic and GeoLife \cite{GeoLife} trajectory datasets.
\end{itemize}

The remainder of the paper is organized as follows. We present the problem statement in Section \ref{sec:ProbStat} where we also introduce privacy and utility metrics. In Section \ref{sec:PUgeneral}, we introduce simplified data release mechanisms for the time-series data PUT problem. In Section \ref{sec:MDPform}, we reformulate the problem as an MDP and propose a numerical evaluation approach utilizing advantage actor-critic deep RL. In Section \ref{sec:SimRes}, we apply the proposed solution to the location trace privacy problem, and compare the performance of the proposed location release strategy with a myopic policy numerically. Finally, we conclude our work in Section \ref{sec:Conc}. 


\vspace{-0.3cm}
\section{Problem Statement}
\label{sec:ProbStat}

\begin{small}
\begin{table}[pt]
    \centering
    \begin{tabular}{c l} 
     \hline
     Notation & Definition \\
     \hline
     $\mathcal{W}$               & Time-series data set              \\ 
     $n$                         & Time-series data length                       \\
     $X_t, Y_t$                  & Random variables representing the user's true   \\
                                 & and distorted data at time $t$             \\
     $p_{x_1}$                   & Probability distribution of the true data at $t=1$ \\
     $q_x(.|.)$                  & Markov transition of user data\\
     $ \mathcal{Q}_x $           & Markov transition matrix of transition probabilities \\
     $q(.|.)$                    & Conditional probability distribution, (policy) \\
     $\mathcal{Q}_H$ &  Probability space of history dependent policies \\  
     $\mathcal{Q}_S, \mathcal{Q'}$ &  Probability space of simplified policies under first-order\\ 
                                   & and $m$-th order Markov assumptions \\
     \hline
    \end{tabular}
    \caption{Notation summary}
    \label{tb:symbols}
\end{table}
\end{small}

We consider a time-series $\{X_t\}_{t\geq1}$, taking values from a finite discrete set $\mathcal{W}$. The user shares $\{X_t\}$ with an SP to gain utility through some online service. 
We assume that the user's true data sequence $\{X_t\}_{t\geq 1}$ follows a first-order time-homogeneous Markov chain with transition probabilities $q_x(x_{t+1}|x_t)$, and initial probability distribution $p_{x_1}$. While the first-order Markov structure assumed for the true data may seem restrictive, we will show that our solution techniques generalize to higher-order Markov chains, albeit with increased complexity in the numerical solutions. In the literature, Markov structure is a common assumption for time-series data, and it is proved to be a reasonable assumption for location trajectories \cite{MarkovProofLocation}, smart meter measurements \cite{MarkovProofSM} and financial data \cite{MarkovProofFinance} due to the history dependent behavior of these time-series.

Instead of sharing its true data at time $t$, the user shares a distorted version of her current data, denoted by $Y_t \in \mathcal{W}$. The released data at time $t$, $Y_t$, does not depend on future data samples; i.e., for any $1 < t< n$, $Y_t \rightarrow (X^t, Y^{t-1}) \rightarrow (X_{t+1}^n, Y_{t+1}^n)$ form a Markov chain, where we have denoted the sequence $(X_{t+1}, \dots, X_n)$ by $X_{t+1}^n$, and the sequence $(X_1, \dots, X_t)$ by $X^t$. The notations which have been used throughout the paper are listed in the Table \ref{tb:symbols}.

\begin{figure}[pb]
\centering
\includegraphics[width=6cm]{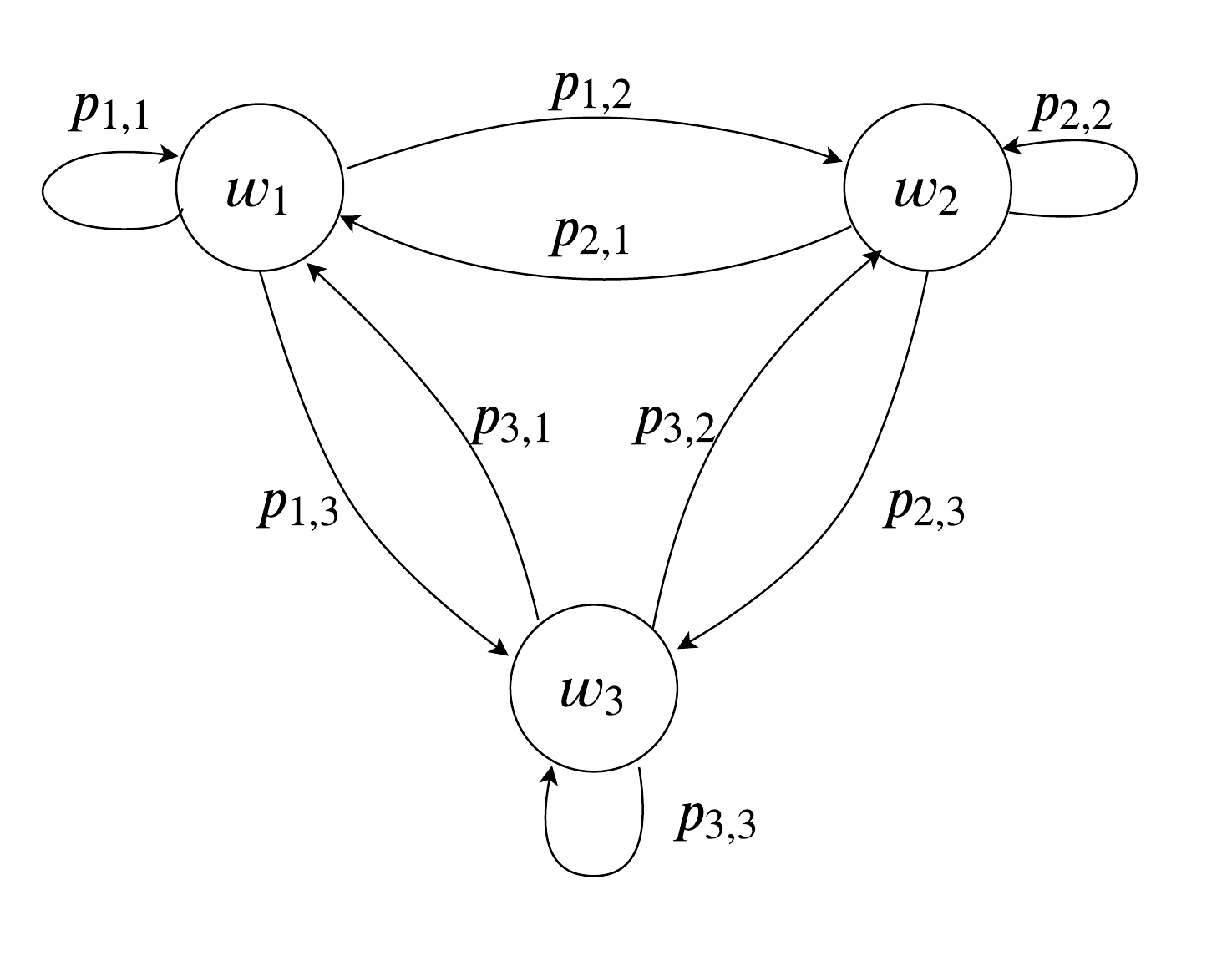}
\caption{Markov chain example for the true data generation.} 
\label{fig:MCgen}
\end{figure}

For a better understanding of the user's private time-series data generation process, a simple Markov chain with state space $\mathcal{W}=\{w_1,w_2,w_3\}$ and state transition probabilities $p_{i,j}$ for $(i,j)\in\{1,2,3\}$ are presented in Fig.\ref{fig:MCgen}. The sensitive data $X_t$ takes the values $\{w_1,w_2,w_3\}$ according to the state transition probabilities. The user becomes aware of $X_t$ in an online manner and releases a distorted version $Y_t \in \{w_1,w_2,w_3\}$, following her privacy-preserving strategy.

\vspace{-0.3cm}
\subsection{Privacy and Utility Measures}

Mutual information can be written as the reduction in the uncertainty of a random variable (r.v.) due to the knowledge of another r.v., i.e., $I(X_t;Y_t)=H(X_t)-H(X_t|Y_t)$, where $H(X_t|Y_t)$ is the conditional entropy. In information theoretic time-series data privacy framework, we assume the strongest model for the malicious third party. That is, both the user and the SP are assumed to have complete statistical knowledge of the user's data as well as her data release mechanism; that is, the transition probabilities of the Markov chain generating the true data sequence and the potentially stochastic mechanism that generates $Y_t$ depending on the history.  
Then, we quantify the privacy by the information leaked to the untrusted SP measured by the mutual information between the true and released data sequences. Accordingly, the information leakage of the user's data release strategy for a time period $n$ is given by
\begin{align}
\hspace{-0.23cm}    I(X^n;Y^n) = \sum\limits_{t=1}^{n}I(X^n;Y_t|Y^{t-1}) = \sum\limits_{t=1}^{n}I(X^t;Y_t|Y^{t-1}), \label{eq:2}
\end{align}
where the first equality follows from the chain rule of mutual information, while the second from the Markov chain $Y^{t}\rightarrow (X_t,Y^{t-1}) \rightarrow X_{t+1}^n$.

Even though a malicious third party can obtain the statistics of the user's data release strategy over an infinite time horizon, i.e., $n \rightarrow \infty$, he cannot infer the realizations of the private information due to the privacy measure based on uncertainty. Since information theoretic metrics are independent of the attack's behavior and computational capabilities, they are preferable as privacy measures. 

In the time-series data privacy problem, we want to minimize the information leakage to the SP. However, as we apply more distortion to the true data sequence for privacy, the more utility is lost due to increased deviation from the original sequence. That is, releasing distorted data reduces the utility received from the SP, and the distortion applied by the user should be limited to a certain level. Therefore, our main purpose is to characterize the trade-off between the privacy and utility. The distortion between the true data sample $X_t$ and the released version $Y_t$ is measured by a distortion measure $d(X_t,Y_t)$ specified based on the underlying application (e.g., Manhattan distance or Euclidean distance), where $d(X_t,Y_t)<\infty, \forall X_t, Y_t \in \mathcal{W}$.

Our main goal is to minimize the information leakage rate to the SP while satisfying the distortion constraint for utility. Throughout the paper, we consider two different constraints on the distortion introduced by PDRP, namely an \textit{instantaneous distortion constraint} and an \textit{average distortion constraint}. The infinite-horizon optimization problem can be written as:
\begin{align}
    \lim_{n \rightarrow \infty} \min_{\substack{\{q_t(y_t|x^t,y^{t-1}):\\d(X_t,Y_t)\leq \bar{D}\}_{t=1}^{n}}} \hspace{0.2cm}  \frac{1}{n}\sum_{t=1}^{n}I^{\boldsymbol{q}}(X^t;Y_t|Y^{t-1}) \label{eq:objectiveIC}
\end{align}
under the instantaneous distortion constraint $\hat{D}$, and as
\begin{align}
    \lim_{n \rightarrow \infty} \min_{\substack{q_t(y_t|x^t,y^{t-1}):\\\mathbb{E}\big[\frac{1}{n}\sum\limits_{t=1}^{n}d(X_t,Y_t)\big]\leq \hat{D}}} \hspace{0.2cm}  \frac{1}{n}\sum_{t=1}^{n}I^{\boldsymbol{q}}(X^t;Y_t|Y^{t-1}) \label{eq:objectiveAC}
\end{align}
under the average distortion constraint $\bar{D}$, where $x_t$ and $y_t$ represent the realizations of $X_t$ and $Y_t$, $\boldsymbol{q}=\{q_t(y_t|x^t,y^{t-1})\}_{t=1}^n$ is a conditional probability distribution which represents the user's randomized \textit{data release policy} at time $t$. The randomness stems from both the Markov process generating the true data sequence, and the random release mechanism $q_t(y_t|x^t,y^{t-1})$.  The mutual information induced by policy $q_t(y_t|x^t,y^{t-1}) \in \boldsymbol{q}$ is calculated using the joint probability distribution
\begin{align}
    P^{\boldsymbol{q}}(& X^n=x^n,Y^n =y^n) \nonumber \\
    &= p_{x_1}q_1(y_1|x_1)\prod_{t=2}^n \big [ q_x(x_t|x_{t-1})q_t(y_t|x^t,y^{t-1}) \big ]. \label{eq:jointH}
\end{align}

In the next section, we characterize the structure of the optimal data release policy, and using this structure we recast the problem as an MDP, and finally evaluate the optimal trade-off numerically using advantage actor-critic deep RL.

\vspace{-0.5cm}
\section{PUT for Time-Series Data Sharing}
\label{sec:PUgeneral}
In this section, we analyze the optimal PUT achievable by a privacy-aware time-series data release mechanism under the notion of mutual information minimization with both instantaneous and average distortion constraints. Moreover, we propose simplified PDRPs that still preserve optimality.

By the definition of mutual information, the objectives (\ref{eq:objectiveIC}) and (\ref{eq:objectiveAC}) depend on the entire history of $X$ and $Y$. Therefore, the user must follow a history-dependent PDRP $q_t^h(y_t|x^t,y^{t-1})$, where the feasible set $\mathcal{Q}_H$ consists of policies that satisfy $\sum_{y_t \in \mathcal{W}}q_t^h(y_t|x^t,y^{t-1})=1$. As a result of strong history dependence, computational complexity of the minimization problem increases exponentially with the length of the data sequence. To tackle this problem, we introduce a class of simplified policies, and prove that they do not cause any loss of optimality in the PUT.

\vspace{-0.5cm}
\subsection{Simplified PDRPs}
In this section we introduce a set of policies $\mathcal{Q}_S\subseteq \mathcal{Q}_H$ of the form $q_t^{s}(y_t|x_t,x_{t-1},y^{t-1})$, which samples the distorted data only by considering the true data in the last two time instances and the entire released data history. Hence, the joint distribution (\ref{eq:jointH}) induced by $\boldsymbol{q}_s \in \mathcal{Q}_S$, where $\boldsymbol{q}_s=\{q_t^{s}(y_t|x_t,x_{t-1},y^{t-1})\}_{t=1}^n$ can be written as
\begin{align}
    &P^{\boldsymbol{q}_s}(X^n= x^n ,Y^n =y^n) \nonumber \\
    &= p_{x_1}q_1^{s}(y_1|x_1) \prod_{t=2}^n \big [ q_x(x_t|x_{t-1})q_t^{s}(y_t|x_t,x_{t-1},y^{t-1}) \big ].\label{eq:JointProb}
\end{align}

Next, we show that considering PDRPs in set $\mathcal{Q}_S$ is without loss of optimality.

\begin{theorem}
In both minimization problems (\ref{eq:objectiveIC}) and (\ref{eq:objectiveAC}), there is no loss of optimality in restricting the PDRPs to the set of policies $\boldsymbol{q}_s \in \mathcal{Q}_S$. Furthermore, information leakage induced by any $\boldsymbol{q}_s \in \mathcal{Q}_S$ can be written as:
\begin{align}
    &I^{\boldsymbol{q}_s}(X^n,Y^n) = \sum\limits_{t=1}^{n}I^{\boldsymbol{q}_s}(X_t,X_{t-1};Y_t|Y^{t-1}) \label{eq:ObjectiveMI}\\
    &=\sum\limits_{t=1}^{n} \hspace{-0.3cm} \sum \limits_{\substack{y^t \in \mathcal{W}^t \\ x_t,x_{t-1} \in \mathcal{W}}}  \hspace{-0.6cm}
    P^{\boldsymbol{q}_s}(x_t,x_{t-1},y^t)\log \frac{q_t^s(y_t|x_t,x_{t-1},y^{t-1})}{P^{\boldsymbol{q}_s}(y_t|y^{t-1})},
    \label{eq:theo1MI}
\end{align}
and the average distortion induced by any $\boldsymbol{q}_s \in \mathcal{Q}_S$ can be written as:
\begin{align}
    \mathbb{E}^{\boldsymbol{q}_s}&\Big[\frac{1}{n}\sum\limits_{t=1}^{n}d(X_t,Y_t)\Big]=\frac{1}{n}\sum\limits_{t=1}^n\mathbb{E}^{\boldsymbol{q}_s}[d(X_t,Y_t)] \label{eq:LinExp} \\
    &=\frac{1}{n}\sum\limits_{t=1}^{n} \sum \limits_{\substack{y^t \in \mathcal{W}^t, \\ x_t,x_{t-1} \in \mathcal{W}}}P^{\boldsymbol{q}_s}(x_t,x_{t-1},y^t)d(x_t,y_t),
    \label{eq:theo1Dist}
\end{align}
where the first equation comes from the linearity of expectation.
\label{prop:1}
\end{theorem}

See Appendix \ref{apx:1} for the proof of Theorem \ref{prop:1}.

\begin{remark}
Although the proof of Theorem \ref{prop:1} assumes that the true data sequence is a first-order Markov chain, it is possible to generalize it to higher-order Markov chains, i.e., $q_x(X_t|X^{t-1})=q_x(X_t|X^{t-1}_{t-m})$ for order $m$. Let $\boldsymbol{Q}_S^m \subseteq \boldsymbol{Q}_H$ denote the set of policies $\boldsymbol{q}'$ 
\begin{align}
    q_t'(y_t|x^t_{t-m},y^{t-1})  =   P^{\boldsymbol{q}'}_{Y_t|X^t_{t-m},Y^{t-1}}(y_t|x^t_{t-m},y^{t-1}).
\end{align}
Then the following theorem holds.
\end{remark}

\begin{theorem}
\label{thm:mthOrder}
If the true data sequence $\{X_t\}$ is a Markov chain of order $m$, then there is no loss of optimally in using a PDRP from the set $\boldsymbol{Q}_S^m$. Moreover, information leakage induced by $\boldsymbol{q}' \in \boldsymbol{Q}_S^m$ can be written as:
\begin{align}
    I^{\boldsymbol{q}'}(X^n,Y^n) = \sum\limits_{t=1}^{n}I^{\boldsymbol{q}'}(X^t_{t-m+1};Y_t|Y^{t-1}), 
\end{align}
and the average distortion induced by any $\boldsymbol{q}' \in \mathcal{Q}^m_S$ can be written as:
\begin{align}
    \mathbb{E}^{\boldsymbol{q}'}\Big[\frac{1}{n}\sum\limits_{t=1}^{n}d(X_t,Y_t)\Big] =\sum\limits_{t=1}^{n} \hspace{-0.5cm} \sum \limits_{\substack{y^t \in \mathcal{W}^t, \\ x^t_{t-m+1} \in \mathcal{W}^{m-1}}} \hspace{-0.7cm} P^{\boldsymbol{q}'}(x^t_{t-m+1},y^t)d(x_t,y_t).
    \label{eq:theo2Dist}
\end{align}
\end{theorem}


\begin{figure}[pt]
\centering
\includegraphics[width=8.8cm,height=2.6cm]{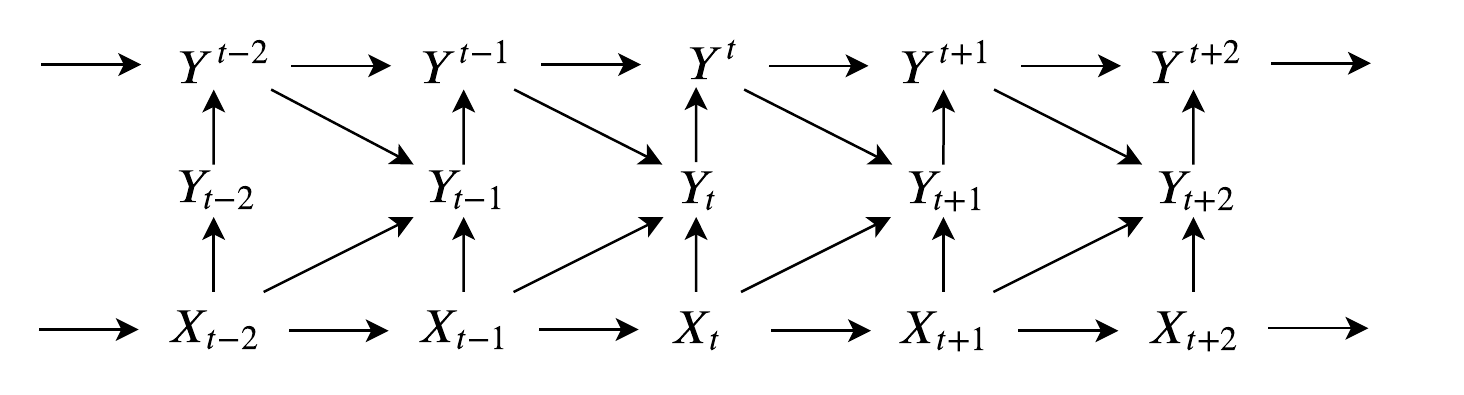}
\caption{Markov chain induced by the simplified PDRP.} 
\label{fig:MC}
\end{figure}

Then the simplified PDRP followed by the user is illustrated by the Markov chain in Fig. \ref{fig:MC}, where $Y^t$ denotes the released data history, i.e., $\{Y_1, \dots, Y_t\}$. That is, the user samples the distorted data, $Y_t$, at time $t$ following $q_t^s(y_t|x_t,x_{t-1},y^{t-1})$ by considering the current and previous true data, $(X_t,X_{t-1})$, and the released data history, $Y^{t-1}$.

\vspace{-0.3cm}
\subsection{Online PDRP with an Instantaneous Distortion Constraint}
\label{subsec:PU_IDC}

As we have stated earlier, we are assuming that the utility gained by the user by sharing its private data diminishes as the distortion between the true data sequence and the released version increases, under the specified distortion measure. Therefore, the utility requirements of the user imposes distortion constraints on the PDPR. Here, we assume that the user would like to guarantee a minimum utility level at each time instant, which, in turn, imposes an instantaneous constraint on the distortion between the true data sample $X_t$ and the released version $Y_t$ at each time instance, i.e., $d(x_t,y_t) \leq \hat{D}$, $\forall t$.

Accordingly, given $(X_t,X_{t-1},Y^{t-1})=(x_t,x_{t-1},y^{t-1})$, the set of feasible simplified PDRPs satisfying an instantaneous distortion constraint is $\boldsymbol{q}_s^{I} \in \boldsymbol{Q}_S^{I}$, and the set of the released data samples induced by $\boldsymbol{q}_s^{I}$ is given by
\begin{align}
    \mathcal{Y}^{\boldsymbol{q}_s^{I}}(x^t_{t-1},y^{t-1}) := \Big \{  y_t \in \mathcal{W}:d(x_t,y_t) \leq \hat{D}  \Big \}. \label{eq:FeasYtIC}
\end{align}
Furthermore, we require $\boldsymbol{q}_s^{I}$ to satisfy 
\begin{align}
    \sum \limits_{y_t \in \mathcal{Y}^{\boldsymbol{q}_s^{I}}(x^t_{t-1},y^{t-1})}  \hspace{-0.7cm}  q_s^{I}(y_t|x^t_{t-1},y^{t-1})=1.
\end{align}
The objective of the PUT for online PDRP with an instantaneous distortion constraints (PDRP-IDC) can be rewritten as
\begin{align}
\hspace{-0.3cm} \min_{q_s^{I}(y_t|x^t_{t-1},y^{t-1})} \hspace{0.1cm} \frac{1}{n} \sum_{t=1}^{n}I^{\boldsymbol{q}_s^{I}}(X_t,X_{t-1};Y_t|Y^{t-1}). \label{eq:IDCobj}
\end{align}

\subsection{Online PDRP with an Average Distortion Constraint}
\label{subsec:PU_ADC}

Alternatively, the user may want to limit only the average distortion applied to the true-data sequence. That is, the utility loss averaged over the time horizon $n$ is denoted by $D(x^n;y^n)=\mathbb{E}[\frac{1}{n} \sum_{t=1}^n d(x_t,y_t)]$. The feasible set of simplified PDRPs with an average distortion constraint is $\boldsymbol{q}_s^{A} \in \boldsymbol{Q}_S^{A}$, and the feasible set of the released $Y_t$ induced by $\boldsymbol{q}_s^{A}$ is given by 
\begin{align}
    \mathcal{Y}^{\boldsymbol{q}_s^{A}}(x^t_{t-1},y^{t-1}) := \Big \{  y_t \in \mathcal{W}:D(x^n,y^n) \leq \bar{D}  \Big \},
\end{align}
where the constraint follows from the linearity of expectation, i.e., $D(x^n;y^n)=\frac{1}{n}\sum_{t=1}^n\mathbb{E}^{\boldsymbol{q}_s^A}[d(x_t,y_t)]$, and the expectation is taken over the joint probabilities of $x_t$ and $y_t$.
Similarly to (\ref{eq:FeasYtIC}), $\boldsymbol{q}_s^A$ is required to satisfy 
\begin{align}
    \sum \limits_{y_t \in \mathcal{Y}^{\boldsymbol{q}_s^{A}}(x^t_{t-1},y^{t-1})}  \hspace{-0.7cm}  q_s^{A}(y_t|x^t_{t-1},y^{t-1})=1.
\end{align}
Hence, the objective of the problem for online PDRP with an average distortion constraint (PDRP-ADC) can be written as:
\begin{align}
    \min_{q_s^A(y_t|x^t_{t-1},y^{t-1})} \hspace{0.1cm} \frac{1}{n} \sum_{t=1}^{n}I^{\boldsymbol{q}_s^A}(X_t,X_{t-1};Y_t|Y^{t-1}). \label{eq:ADCobj}
\end{align}

Minimization of the mutual information subject to a distortion constraint can be converted into an unconstrained minimization problem using Lagrange multipliers. Since the distortion constraint induced by the simplified PDRP is memoryless, we can integrate it into the additive mutual information objective easily. Hence, the unconstrained minimization problem for time-series data release PUT can be rewritten as
\begin{align}
    \min_{\boldsymbol{q}_s \in \boldsymbol{Q}_s} \hspace{-0.1cm} \frac{1}{n} \sum_{t=1}^{n} \hspace{-0.1cm}  \big [ I^{\boldsymbol{q}_s}(X_t,X_{t-1};Y_t|Y^{t-1}) \hspace{-0.05cm} + \hspace{-0.05cm} \lambda (\mathbb{E}^{\boldsymbol{q}_s}[d(x_t,y_t)] \hspace{-0.1cm} - \hspace{-0.1cm} \bar{D}) \big ], \label{eq:UnconsObjective}
\end{align}
where $\lambda$ is the Lagrangian multiplier, and determines the operating point on the trade-off curve, i.e., it represents where the gradients of the mutual information and the distortion constraint point in the same direction. When $\lambda=0$, the user releases data samples which only minimize the information leakage. On the other hand, as $\lambda \rightarrow \infty$, the released data minimizes only distortion constraint rather than information leakage, which results in full information leakage.

In the following section, we present the MDP formulation of the problem for both PDRPs and the evaluation method utilized by advantage actor-critic RL.

\section{MDP Formulation}
\label{sec:MDPform}
Markovity of the user's true data sequence and the additive objective functions in both (\ref{eq:IDCobj}) and (\ref{eq:UnconsObjective}) allow us to represent the problem as an MDP with state $X_t$. However, the information leakage at time $t$ depends on $Y^{t-1}$, resulting in a growing state space in time. Therefore, for a given policy $\boldsymbol{q}_s$ and any realization $y^{t-1}$ of $Y^{t-1}$, we define a belief state $\beta_t \in \mathcal{P}_X$ as a probability distribution over the state space:

\begin{align}
    \beta_t(x_{t-1})=P^{\boldsymbol{q}_s}(X_{t-1}=x_{t-1}|Y^{t-1}=y^{t-1}).
\end{align}

This represents the SP's belief on the true data sample at the beginning of time instance $t$, i.e., after receiving the distorted-data $y_{t-1}$. The actions are defined as probability distributions with which the user samples the released value $Y_t$ at time $t$ and determined by the randomized PDRPs. The user's action induced by a policy $\boldsymbol{q}_s$ can be denoted by $a_t(y_t|x_t,x_{t-1})=P^{\boldsymbol{q}_s}(Y_t=y_t|X_t=x_t,X_{t-1},\beta_t)$. At each time $t$, the SP updates its belief on the true data sample $\beta_{t+1}(x_t)$, after observing its distorted version $y_t$ by

\begin{align}
    & \beta_{t+1}(x_{t}) =\frac{p(x_{t},y_t|y^{t-1})}{p(y_t|y^{t-1})}= \frac{\sum_{x_{t-1}}p(x_{t},x_{t-1},y_t|y^{t-1})}{\sum_{x_t,x_{t-1}}p(x_{t},x_{t-1},y_t|y^{t-1})} \nonumber \\
    & = \frac{\sum_{x_{t-1}}p(x_{t}|x_{t-1})q_t^s(y_t|x_t,x_{t-1},y^{t-1})p(x_{t-1}|y^{t-1})}{\sum_{x_t,x_{t-1}}p(x_{t}|x_{t-1})q_t^s(y_t|x_t,x_{t-1},y^{t-1})p(x_{t-1}|y^{t-1})} \nonumber\\
    &= \frac{\sum_{x_{t-1}}q_x(x_{t}|x_{t-1})a(y_t|x_t,x_{t-1})\beta_t(x_{t-1})}{\sum_{x_t,x_{t-1}}q_x(x_{t}|x_{t-1})a(y_t|x_t,x_{t-1})\beta_t(x_{t-1})}.
    \label{eq:BeliefUpdate}
\end{align}
We define the per-step information leakage of the user due to taking the action $a_t(y_t|x_t,x_{t-1})$ at time $t$ as,
\begin{align} 
\label{eq:PerStepObjective}
 l_t(x_t,x_{t-1},a_t,y^t;\boldsymbol{q}_s) := \log \frac{a_t(y_t|x_t,x_{t-1})}{P^{\boldsymbol{q}_s}(y_t|y^{t-1})}.
\end{align}

\sloppy
The expectation of $n$-step sum of (\ref{eq:PerStepObjective}) over the joint probability $P^{\boldsymbol{q}_s}(X_t,X_{t-1},Y^t)$ is equal to the mutual information expression in the original problem (\ref{eq:ObjectiveMI}). Therefore, given the belief and action probabilities, average information leakage at time $t$ can be formulated as,
\begin{align}
\mathbb{E}^{\boldsymbol{q}_s}[l_t(x^t_{t-1},&a_t,y^t)]  \hspace{-0.05cm} = \hspace{-0.8cm}
\sum_{ \substack{x_t,x_{t-1},y_t \in \mathcal{W}}}  \hspace{-0.7cm}   \beta_t(x_{t-1})a_t(y_t|x_t,x_{t-1})q_x(x_{t}|x_{t-1}) \nonumber \\  
& \times \log \frac{a_t(y_t|x_t,x_{t-1})}{\hspace{-0.5cm}\sum\limits_{\substack{\hat{x}_t,\hat{x}_{t-1} \in \mathcal{W}}} \hspace{-0.5cm}  \beta_t(\hat{x}_{t-1})a_t(y_t|\hat{x}_t,\hat{x}_{t-1})q_x(\hat{x}_{t}|\hat{x}_{t-1})}\nonumber \\
&:=\mathcal{L}(\beta_t,a_t).
\label{eq:AvgLeakage}
\end{align}

We can recast the PDRP-IDC problem in (\ref{eq:IDCobj}) as a continuous state and action space MDP. The actions satisfying the instantaneous distortion constraint are denoted by $a^{\text{IDC}}_t(y_t|x_t,x_{t-1})$ and induced by the simplified PDRP $q_s^I(y_t|x^t_{t-1},y^{t-1})$.
The solution of the MDP for PDRP-IDC problem relies on minimizing the objective
\begin{align}
    \mathcal{C}_{\text{IDC}}(\beta_t,a^{\text{IDC}}_t):=\mathcal{L}(\beta_t,a^{\text{IDC}}_t),
\end{align}
where $\mathcal{L}(\beta_t,a^{\text{IDC}}_t)$ is the average information leakage obtained by taking the actions $a^{\text{IDC}}_t(y_t|x_t,x_{t-1})$, at each time step $t$.

We remark that the representation of average distortion in terms of belief and action probabilities is straightforward due to its additive form. Similarly to (\ref{eq:AvgLeakage}), average distortion for PDRP-ADC at time $t$ can be written as,
\begin{align}
    \mathbb{E}^{\boldsymbol{q}_s}[d(x_t,y_t)] \hspace{-0.05cm} &= \hspace{-0.8cm}  \sum_{x_t,x_{t-1},y_t\in \mathcal{W}} \hspace{-0.70cm}  \beta_t(x_{t-1} \hspace{-0.05cm} )a_t(y_t|x_t,x_{t-1} \hspace{-0.05cm} )q_x( \hspace{-0.05cm} x_t|x_{t-1} \hspace{-0.05cm} )d(x_t,y_t \hspace{-0.05cm} ) \nonumber \\
    &:= \mathcal{D}(\beta_t,a_t),
\end{align}
where there is no restriction on how the actions are chosen, i.e., $y_t \in \mathcal{W}$.
Hence, we can recast the PDRP-ADC problem in (\ref{eq:UnconsObjective}) as a continuous state and action space MDP with a per-step cost function given by
\begin{align}
    \mathcal{C}_{\text{ADC}}(\beta_t,a_t):=\mathcal{L}(\beta_t,a_t)+\lambda (\mathcal{D}(\beta_t,a_t)-\hat{D}). \label{eq:MDPCost}
\end{align}

Finding optimal policies for continuous state and action space MDPs is a PSPACE-hard problem \cite{PSPACEhard}.
In practice, they can be solved by various finite-state MDP evaluation methods, e.g., value iteration, policy iteration and gradient-based methods. These are based on the discretization of the continuous belief states to obtain a finite state MDP \cite{Tamas}. While finer discretization of the belief reduces the loss from the optimal solution, it causes an increase in the dimension of the state space; hence, in the complexity of the problem. To overcome the complexity limitation, we will employ a deep learning based method as a tool to numerically solve our continuous state and action space MDP problem.

\vspace{-0.5cm}
\subsection{Advantage Actor-Critic (A2C) Deep RL}

\begin{figure}[pt]
\centering
\includegraphics[width=7.5cm]{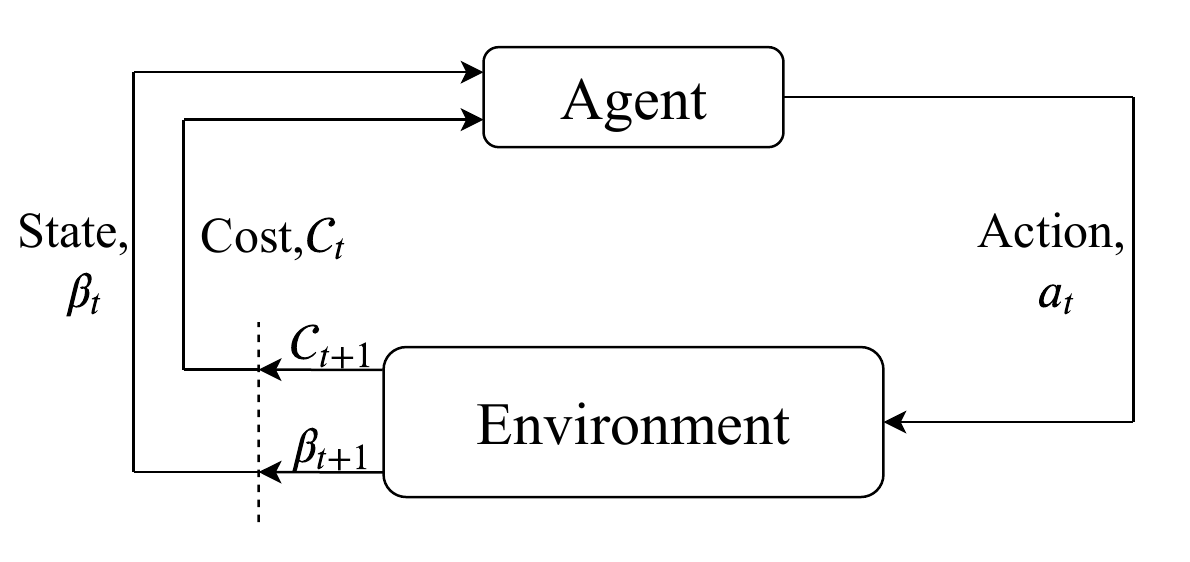}
\caption{RL for a known model.} 
\label{fig:RL}
\end{figure}

In this section, we simply use $\mathcal{C}(\beta_t,a_t)$ and $a_t(y_t|x_t,x_{t-1})$ to represent the MDP cost and action pair of both PDRP-IDC and PDRP-ADC, respectively. Integration of the solution into the instantaneous and average distortion constrained cases is straightforward.

In RL, an agent discovers the best action to take in a particular state by receiving instantaneous rewards/costs from the environment \cite{SuttonBarto}. On the other hand, in our problem, we have the knowledge of the state transition probabilities and the cost for every state-action pair without the need for interacting with the environment. We use A2C-deep RL as a computational tool to numerically evaluate the optimal PDRP for our continuous state and action space MDP.

To integrate RL framework into our problem, we create an artificial environment which inputs the user's current action, $a_t(y_t|x_t,x_{t-1})$, samples an observation $y_t$, and calculates the next state, $\beta_{t+1}$, using Bayesian belief update (\ref{eq:BeliefUpdate}). Instantaneous cost revealed by the environment is calculated by (\ref{eq:MDPCost}). The user receives the experience tuple $(\beta_t,a_t,y_t,\beta_{t+1},\mathcal{C}_t)$ from the environment, and refines her policy accordingly.
Fig. \ref{fig:RL} illustrates the interaction between the artificial environment and the user, which is represented by the RL agent.
The corresponding Bellman equation induced by policy ${\boldsymbol{q}_s}$ can be written as 
\begin{align}
    V^{\boldsymbol{q}_s}(\beta)+J({\boldsymbol{q}_s})= \min_{a}\Big\{ \mathcal{C}(\beta,a)+V^{\boldsymbol{q}_s}(\beta') \Big\},
    \label{eq:Bellman}
\end{align}
where $V^{\boldsymbol{q}_s}(\beta)$ is the state-value function, $\beta'$ is the updated belief state according to (\ref{eq:BeliefUpdate}), $a$ represents action probability distributions, and $J({\boldsymbol{q}_s})$ is the cost-to-go function, i.e., the expected future cost induced by policy ${\boldsymbol{q}_s}$ \cite{Bertsekas}.

\begin{figure}[pt]
\centering
\hspace{-0.8cm}
\subfloat{}{
\includegraphics[width=7cm]{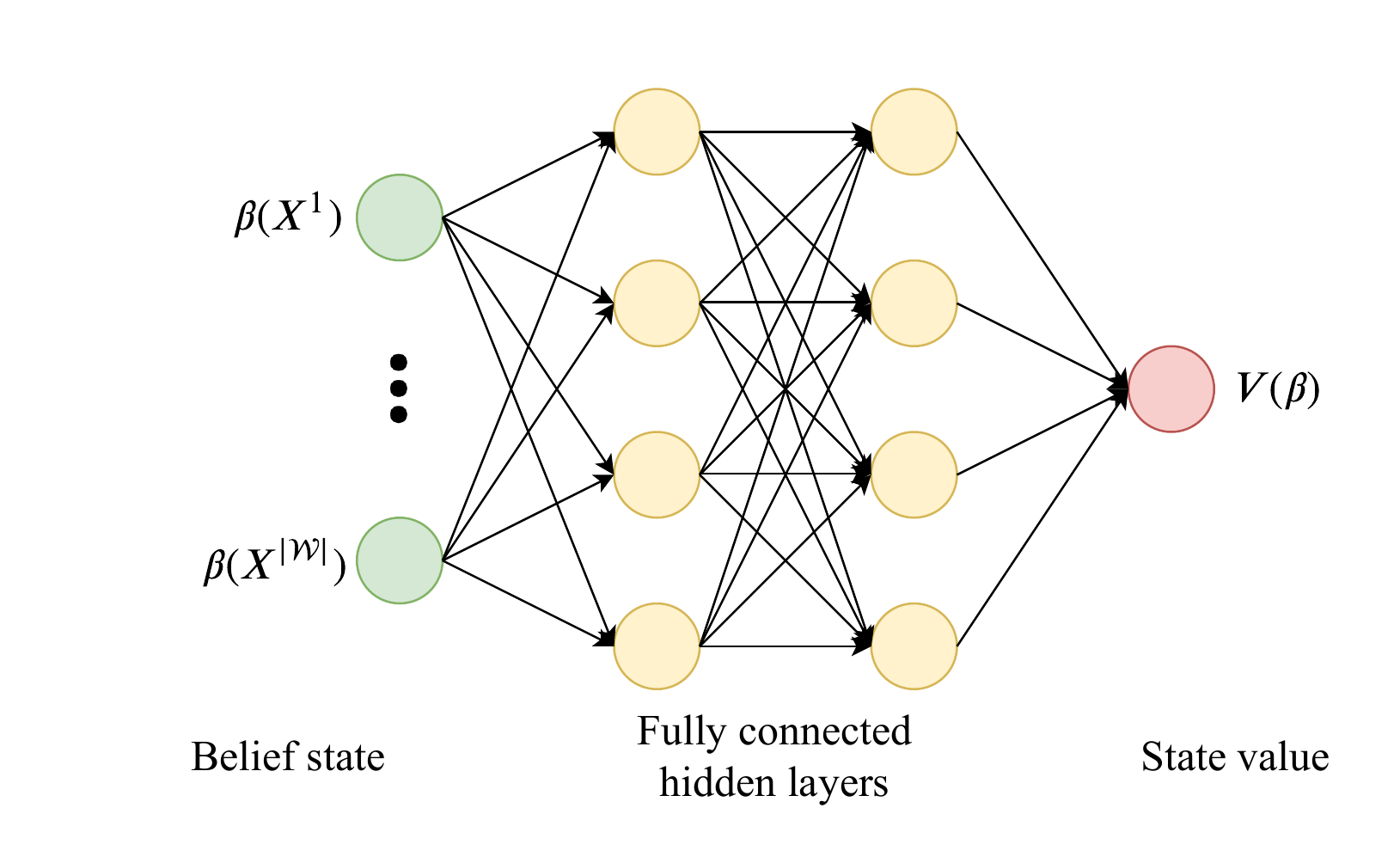}%
\caption*{(a)}
\label{fig:DeepCritic}}
\vfill
\subfloat{}{
\includegraphics[width=7.3cm]{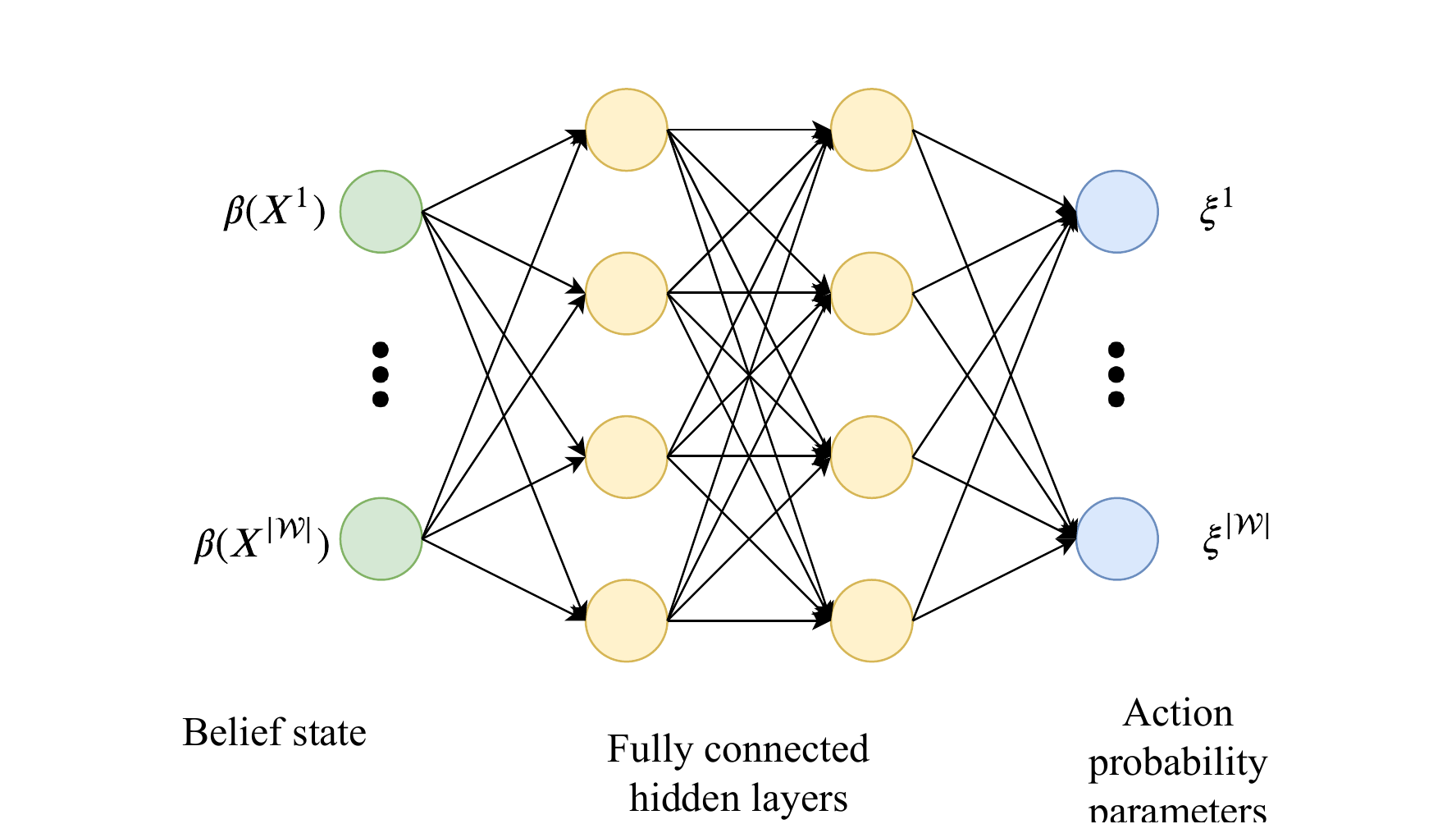}%
\caption*{(b)}
\label{fig:DeepActor}}
\caption{Critic (a) and actor (b) DNN structures.}
\label{fig:DeepNN}
\end{figure}

RL methods can be divided into three groups: value-based, policy-based, and actor-critic \cite{OnACalgs}. Actor-critic methods combine the advantages of value-based (critic-only) and policy-based (actor-only) methods, such as low variance and continuous action producing capability. The actor represents the policy structure, while the critic estimates the value function \cite{SuttonBarto}. In our setting, we parameterize the value function by the parameter vector $\boldsymbol{\theta} \in \Theta$ as $V_{\boldsymbol{\theta}}(\beta)$, and the stochastic policy by $\boldsymbol{\xi} \in \Xi$ as $q_{\boldsymbol{\xi}}$. The difference between the right and the left hand side of (\ref{eq:Bellman}) is called temporal difference (TD) error, which represents the error between the critic's estimate and the target differing by one-step in time \cite{SurveyACRL}. The TD error for the experience tuple $(\beta_t,a_t,y_t,\beta_{t+1},\mathcal{C}_t)$ is estimated as
\begin{align}
    \delta_t=\mathcal{C}_t(\beta_t,a_t)+\gamma V_{\theta_t}(\beta_{t+1})-V_{\theta_t}(\beta_t),
\end{align}
where $\mathcal{C}_t(\beta_t,a_t)+\gamma V_{\theta_t}(\beta_{t+1})$ is called the TD target, and $\gamma$ is a discount factor that we choose very close to $1$ to approximate the Bellman equation in (\ref{eq:Bellman}) for our infinite-horizon average cost MDP. To implement RL in the infinite-horizon problem, we take sample averages over independent and finite data sequences, which are generated by experience tuples at each time $t$ via Monte-Carlo roll-outs.

Instead of using value functions in actor and critic updates, we use advantage function to reduce the variance in policy gradient methods. The advantage can be approximated by TD error. Hence, the critic is updated by gradient descent as:
\begin{align}
    \theta_{t+1}=\theta_t+\eta_t^c \nabla_{\theta}\ell_{c}(\theta_t),
\end{align}
where $\ell_c(\theta_t)=\delta_t^2$ is the critic loss and $\eta_t^c$ is the learning rate of the critic at time $t$. The actor is updated similarly as, 
\begin{align}
    \xi_{t+1}=\xi_t - \eta_t^a \nabla_{\xi}\ell_a(\xi_t),
\end{align}
where $\ell_a(\xi_t)=\ln(q_s(y_t|\beta_t,\xi_t))\delta_t$ is the actor loss and $\eta_t^a$ is the actor's learning rate. This method is called \textit{advantage actor-critic RL}.

\begin{algorithm}[pt]
\SetAlgoLined
 Initialize DNNs with random weights $\xi$ and $\theta$ \\
 Initialize environment $E$\\
 \For{episode=$1,N$}{
  Initialize belief state $\beta_0$\;
  \For{$t=0,n$}{
   Sample action probability vector $a_t \sim Dirichlet(a|\xi)$\ according to the current policy;\\
   Perform action $a_t$ and calculate cost $\mathcal{C}_{\xi_t}$ in $E$;\\
   Sample an observation $y_t$ and calculate next belief state $\beta_{t+1}$ in $E$;\\
   Set TD target $\mathcal{C}_{\xi_t}+\gamma V_{\theta_t}^{\xi}(\beta_{t+1})$;\\
   Minimize the loss $\ell_c(\theta)=\delta^2=(\mathcal{C}_{\xi_t}+\gamma V_{\theta_t}^{\xi}(\beta_{t+1})-V_{\theta_t}^{\xi}(\beta_t))^2$;\\
   Update the critic $\theta \leftarrow \theta + \eta^c \nabla_{\theta}\delta^2$;\\
   Minimize the loss $\ell_a(\xi_t)=\ln(Dirichlet(a|\xi_t))\delta_t$;\\
   Update actor $\xi \leftarrow \xi -\eta^a \nabla_{\xi}\ell_a(\xi_t)$;\\
   Update belief state $\beta_{t+1} \leftarrow \beta_t$ 
   }{
  }
 }
 \caption{A2C-deep RL algorithm for online PDRP} \label{alg:A2CDRL}
\end{algorithm}

In our A2C-deep RL implementation, we represent the actor and critic mechanisms by fully connected feed-forward deep neural networks (DNNs) with two hidden layers as illustrated in Fig. \ref{fig:DeepNN}. The critic DNN takes the current belief state $\beta(\boldsymbol{X})$ of size $|\mathcal{W}|$ as input, where $\boldsymbol{X}$ is the true data sequence vector, and outputs the value of the belief state for the current action probabilities $V_{\theta}^{\xi}(\beta)$. The actor DNN also takes the current belief state $\beta(\boldsymbol{X})$ as input, and outputs the parameters used for determining the action probabilities of the corresponding belief. Hence, the input/output sizes of the critic and actor DNNs are $|\mathcal{W}| \times 1$  and $|\mathcal{W}| \times |\mathcal{W}|$, respectively. Here, the actor DNN output parameters $\{\xi^1, \dots, \xi^{|\mathcal{W}|}\}$ are used to generate a Dirichlet distribution, which represents the action probabilities. The overall A2C-deep RL algorithm for online PDRP is described in Algorithm \ref{alg:A2CDRL}. In the next section, we apply the proposed deep RL solution to a location trace privacy problem.

\vspace{-0.3cm}
\section{Application to Location Trace Privacy}
\label{sec:SimRes}
\begin{figure}[pt]
\centering
\includegraphics[width=5.6cm]{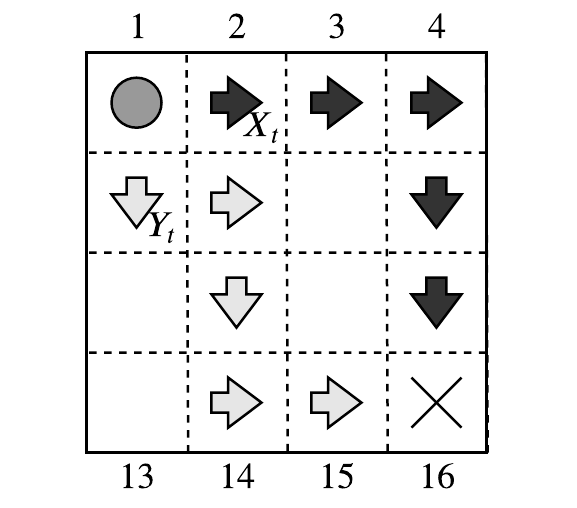}
\caption{True and released user trajectory example for $n=5$. } 
\label{fig:Trajectory}
\end{figure}

In this section, we consider an application of the theoretical framework we have introduced to the location trace privacy problem. We focus on location trace as an example of time-series data. In this scenario, the user shares a distorted version of her trajectory with the SP due to privacy concerns. An example for the user trajectory of length $n=5$ in a grid area is illustrated in Fig. \ref{fig:Trajectory}. While the user's location at time $t=0$ is depicted with a grey circle, the true and released user trajectories over the next $5$ time steps are represented by black and grey arrows, respectively.

\vspace{-0.3cm}
\subsection{Numerical Results for Synthetic Data}
In this section, we evaluate the PUT of the proposed PDRP-ADC and PDRP-IDC methods for synthetic user mobility data. We also compare the PDRP-ADC results with the myopic Markovian location release mechanism proposed in \cite{Ravi}.
For the simulation results presented in the following sections, we train two fully connected feed-forward DNNs, representing the actor and critic networks, respectively, by utilizing ADAM optimizer \cite{ADAM}. Both networks contain two hidden layers of sizes $3000$ with leaky-ReLU activation \cite{LeakyRELU}. We obtain the corresponding PUT by averaging the total information leakage for the specified distortion constraint over a time horizon of $n=300$.

\subsubsection{PDRP-IDC Results}
\label{subsec:SimIDC}

We first consider a simple $4\times4$ grid-world, where $|\mathcal{W}|= 16$ as in Fig. \ref{fig:Trajectory}. The cells are numbered such that the first and the last rows of the grid-world are represented by $\{1,2,3,4\}$ and $\{13,14,15,16\}$, respectively. The user's trajectory forms a first-order Markov chain with a transition probability matrix $\boldsymbol{Q}_x$ of size $|\mathcal{W}| \times |\mathcal{W}|$, whose index $Q_x(i,j)$, $i,j \in \{1, \dots, |\mathcal{W}|\}$, represents the transition probability $q_x(x_{t}=i|x_{t-1}=j)$ from the state $j$ to $i$. The user can start its movement at any square with equal probability, i.e., $p_{x_1}=\frac{1}{16}$. 
Our goal is to obtain the PUT under instantaneous distortion constraints $\hat{D} \in \{1, \dots, 4\}$ with Manhattan distance on the distortion measure between the true position and the reported one.

\begin{figure}[pt]
\centering
\includegraphics[width=8.8cm]{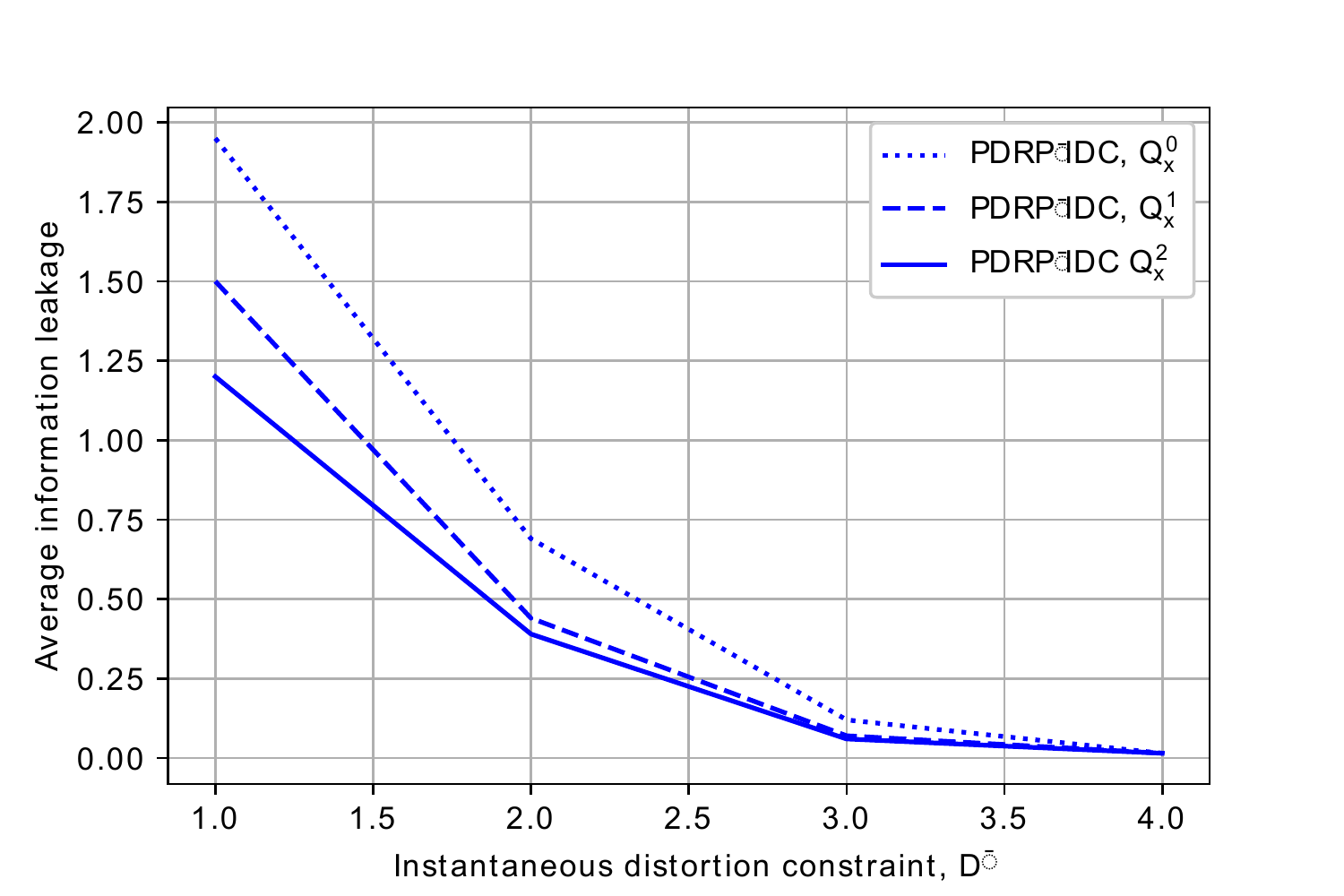}
\caption{Average information leakage as a function of the allowed instantaneous distortion under Manhattan distance as the distortion measure.}
\label{fig:IDC_RD}
\end{figure}

In Fig. \ref{fig:IDC_RD}, PUT curves are obtained for transition probability matrices $\boldsymbol{Q}_x^0$, $\boldsymbol{Q}_x^1$ and $\boldsymbol{Q}_x^2$, each corresponding to a different temporal correlation level. In all the cases, the user can move from any square to any other square in the grid at each step, i.e., $Q_x^m(i,j)>0$, $\forall m, i, j$. While all the transition probabilities are equal to $\frac{1}{|\mathcal{W}|}$ for $\boldsymbol{Q}_x^0$, the probability of the user moving to a nearby square is greater than taking a larger step to a more distant one for $\boldsymbol{Q}_x^1$ and $\boldsymbol{Q}_x^2$.
Moreover, $\boldsymbol{Q}_x^1$ represents a more uniform trajectory, where the agent moves to equidistant cells with equal probability, while with $\boldsymbol{Q}_x^2$ the agent is more likely to follow a certain path, i.e., the random trajectory generated by $\boldsymbol{Q}_x^2$ has lower entropy. The transition probabilities for $\boldsymbol{Q}_x^1$ are given by:
\begin{align}
    q^1_x(x_t|x_{t+1})= \dfrac{ {r_{d(x_t,x_{t+1})}}/{d(x_t,x_{t+1})}  }{ \sum_{x_{t+1}\in \mathcal{W}}  {r_{d(x_t,x_{t+1})}}/{d(x_t,x_{t+1})}  } ,
\end{align}
where $d(x_t,x_{t+1})$ is the Manhattan distance between positions $x_t$ and $x_{t+1}$; $r_{d(x_t,x_{t+1})}$ is a scalar which determines the probability of the user moving from one square to any of the equidistant squares in the next step. Fig. \ref{fig:Sim} is obtained by setting $r_0=1$ and $r_i=7-i$, $i=1,\dots,6$. 

For $\boldsymbol{Q}_x^2$, we set  
\begin{align}
   q^2_x(x_t|x_{t+1})= \frac{ {u(x_t,x_{t+1})}/{d(x_t,x_{t+1})}  }{ \sum_{x_{t+1}\in \mathcal{W}}  {u(x_t,x_{t+1})}/{d(x_t,x_{t+1})}  },  
\end{align}
where, for $x_t \in \{1,2, \dots, 15\}$, we have
\begin{align}
   u(x_t,x_{t+1})\hspace{-0.1cm} = \hspace{-0.1cm} \begin{cases}
    r_1, & \text{for $mod(x_t,4) \neq 0$, $x_{t+1}=x_t+1$},\\
    r_1, & \text{for $mod(x_t,4)=0$, $x_{t+1}=x_t+4$}, \\
    r_0, & \text{otherwise},
 \nonumber \end{cases}
\end{align}
where $mod(.)$ is the modulo operator which finds the remainder after division of $x_t$ by $4$, and $u(16,x_{t+1})=r_0$ for $x_{t+1} \in \{1, \dots, 15\}$, and $u(16,16)=r_1$. As a result, temporal correlations in the location history increase in the order $\boldsymbol{Q}_x^0$, $\boldsymbol{Q}_x^1$, $\boldsymbol{Q}_x^2$.

We train our DNNs for a time horizon of $n=300$ in each episode, and over $5000$ Monte Carlo roll-outs. Fig. \ref{fig:IDC_RD} shows that, information leakage increase in the order $\boldsymbol{Q}_x^2$, $\boldsymbol{Q}_x^1$, $\boldsymbol{Q}_x^0$. As the temporal correlations between the locations on a trace increases, the proposed PDRP-IDC leaks less information since it takes the entire released location history into account.

\subsubsection{PDRP-ADC Results}
\begin{figure}[pt]
\centering
\includegraphics[width=8.8cm]{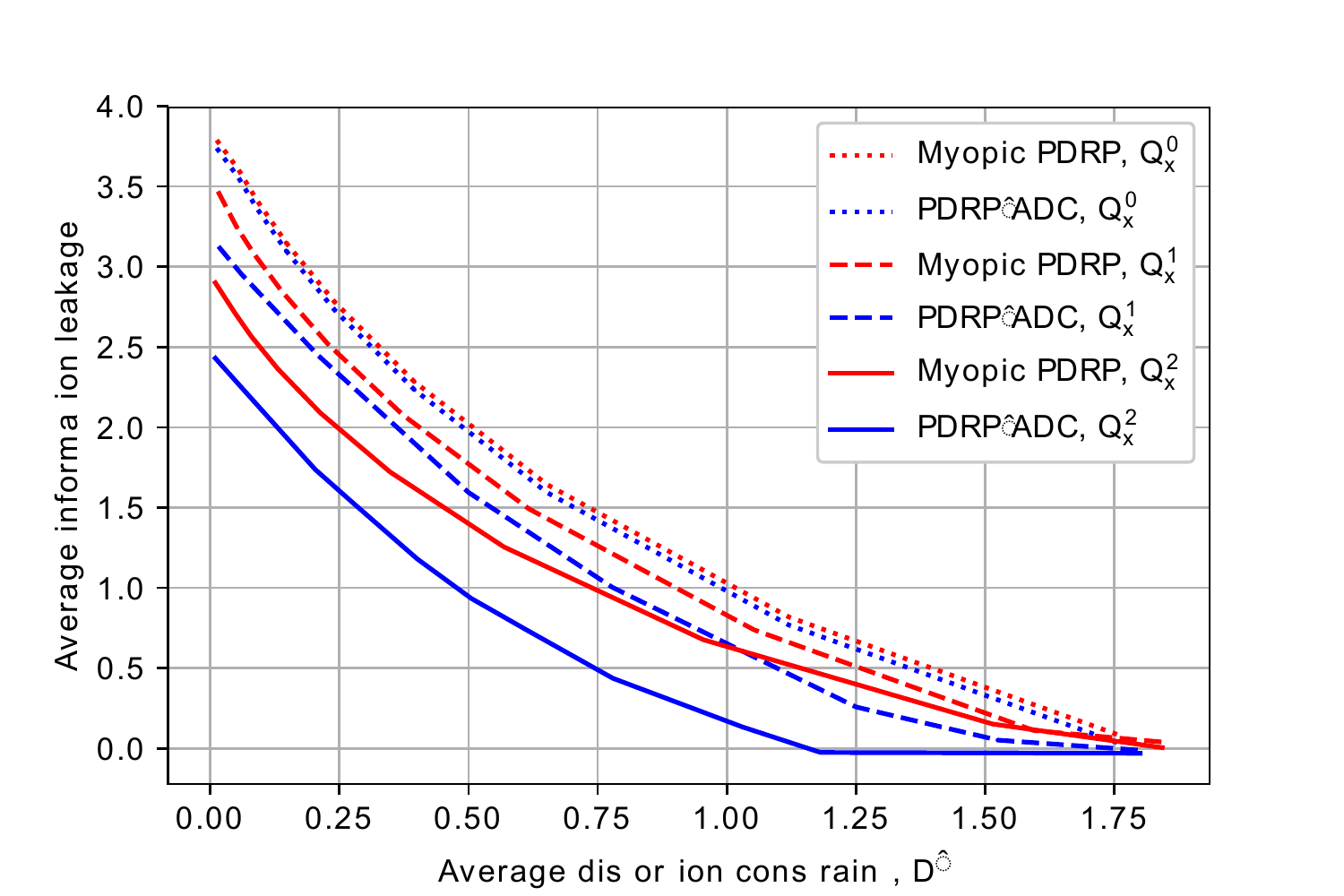}
\caption{Average information leakage as a function of the allowed average distortion under Manhattan distance as the distortion measure.}
\label{fig:Sim}
\end{figure}


Next, we consider the same scenario as before, but evaluate the PUT under an average distortion constraint. We evaluate the performance of the proposed PDRP-ADC and compare the results with the myopic Markovian location release mechanism proposed in \cite{Ravi}. In \cite{Ravi}, an upper bound on the PUT is given by a myopic policy as follows:
\begin{align}
    \sum \limits_{t=1}^{n} \min_{\substack{q(y_t|x_t,x_{t-1},y_{t-1}):\\\mathbb{E}^q[d(x_t,y_t)]\leq \hat{D}}} I^q(X_t,X_{t-1};Y_t|Y_{t-1}). \label{eq:RaviCost}
\end{align}
Exploiting the fact that (\ref{eq:RaviCost}) is similar to the rate-distortion function, Blahut-Arimoto algorithm is used in \cite{Ravi} to minimize the conditional mutual information at each time step. Finite-horizon solution of the objective function (\ref{eq:RaviCost}) is obtained by applying alternating minimization sequentially. In our simulations, we obtained the average information leakage and distortion for this approach by normalizing for $n=300$.

In Fig. \ref{fig:Sim}, PUT curves of the proposed PDRP-ADC and the myopic location release mechanism are obtained for the same environment defined in Section \ref{subsec:SimIDC}. The same transition matrices are used, i.e., $\boldsymbol{Q}_x^0$, $\boldsymbol{Q}_x^1$ and $\boldsymbol{Q}_x^2$ represent increasing temporal correlations in the user's trajectory. The Lagrangian multiplier $\lambda \in [0,20]$ denotes the user's choice for the operating point on the PUT curve. Distortion is again measured by the Manhattan distance. Similarly to Section \ref{subsec:SimIDC}, we train our DNNs for $n=300$ in each episode, and over $5000$ Monte Carlo roll-outs. Fig. \ref{fig:Sim} shows that, for $\boldsymbol{Q}_x^2$ the proposed PDRP-ADC obtained through deep RL leaks much less information than the myopic location release mechanism for the same distortion level, indicating the benefits of considering all the history when taking actions at each time instant. The gain is less for $\boldsymbol{Q}_x^1$, since there is less temporal correlations in the location history compared to $\boldsymbol{Q}_x^2$; and hence, there is less to gain from considering all the history when taking actions. Finally, for $\boldsymbol{Q}_x^0$ the proposed scheme and the myopic policy perform the same, since the user movement with uniform distribution does not have temporal memory; and therefore, taking the history into account does not help.

\begin{figure}[pt]
\centering
\includegraphics[width=8.4cm,height=5.8cm]{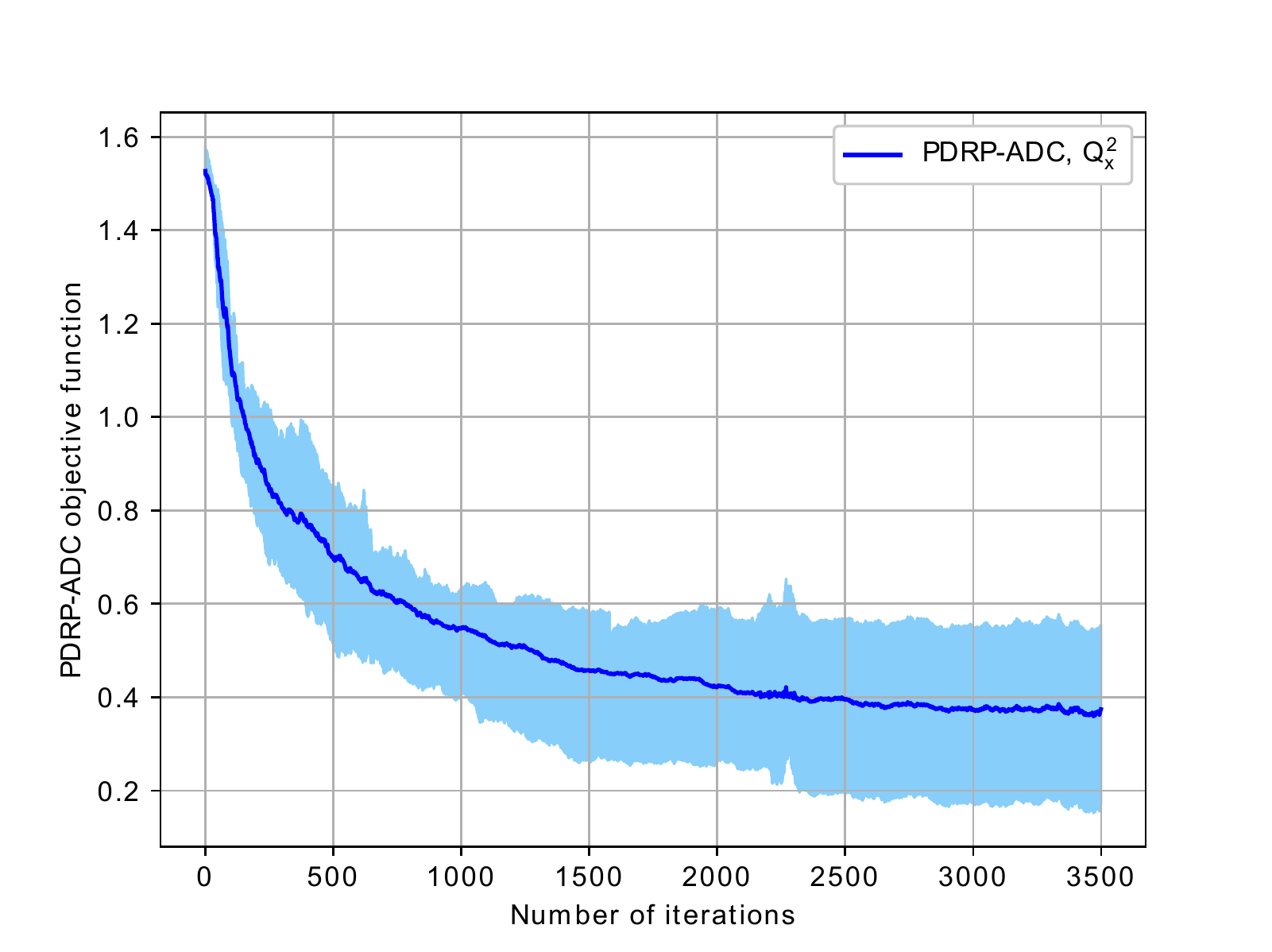}
\caption{Convergence of PDRP-ADC for $\lambda=1$, $\hat{D}=0.8$ and $\mathcal{\boldsymbol{Q}}_x^2$.}
\label{fig:Convergence}
\end{figure}

Fig. \ref{fig:Convergence} shows the convergence behaviour of the A2C-DRL algorithm when evaluating PDRP-ADC's objective function (\ref{eq:UnconsObjective}) for $\mathcal{\boldsymbol{Q}}_x^2$, $\lambda$=$1$, $\hat{D}$=$0.8$. Various realizations of the convergence curve lie in the light blue area, and the dark blue curve represents the average value of these realizations. 
We observe that the convergence typically occurs after about 2500 iterations. On the other hand, we remark that the optimal policy for a stationary environment can be obtained in an offline manner using the available dataset; therefore the convergence time and the number of iterations has no impact on the real-time application of this solution in practice.

We next consider a toy example for PDRP-ADC to visualize the location release strategy for a better understanding. We consider a $2\times3$ grid-world, where the user's trajectory forms a first-order Markov chain with the transition probability matrix $\boldsymbol{Q}_x$, given in Table \ref{tb:6sQx}.
\tabcolsep=0.15cm
\begin{table}[h]
    \centering
    \begin{tabular}{|c| c c c c c c|} 
     \hline
     \diagbox[width=5em]{$x_{t-1}$}{$x_t$} & $1$ & $2$ & $3$ & $4$ & $5$ & $6$ \\  
     \hline
     1 & 0.11 &  0.64 & 0.05 &  0.11 &  0.05 &  0.04 \\
     2 & 0.1  &  0.1  & 0.6  &  0.05 &  0.1  &  0.05  \\
     3 & 0.05 &  0.11 & 0.11 &  0.04 &  0.05 &  0.64  \\
     4 & 0.11 &  0.05 & 0.04 &  0.11 &  0.64 &  0.05  \\
     5 & 0.05 &  0.1  & 0.05 &  0.1  &  0.1  &  0.6  \\ 
     6 & 0.04 &  0.05 & 0.11 &  0.05 &  0.11 &  0.64  \\  
     \hline
    \end{tabular}
    \caption{The transition probability matrix $Q_x$ of the toy example for PDRP-ADC, when $|\mathcal{W}|=6$.}
    \label{tb:6sQx}
\end{table}
We assume that the user can start its movement at any square with equal probability, i.e., $p_{x_1}=\frac{1}{6}$. The Lagrange multiplier is chosen as $\lambda = 3$, and the distortion constraint is $\bar{D}=0.6$.

After training the actor and critic DNNs, we obtain the best action probabilities that minimize the objective function $\mathcal{C}_{\text{ADC}}$ in (\ref{eq:MDPCost}). Given the user trajectory pattern in Table \ref{tb:6sQx}, $\beta=[\frac{1}{6}, \dots, \frac{1}{6}]$ and $\lambda=3$, the action distribution matrix induced by PDRP-ADC is obtained as in Table \ref{tb:6sAt}. It is clear from the table that $Y_t$ is not released according to a deterministic pattern.

\vspace{0.2cm}
\tabcolsep=0.15cm
\begin{table}[t]
    \centering
\begin{tabular}{|c| c c c c c c|} 
  \hline
  \diagbox[width=7em]{$x_t,x_{t-1}$}{$y_t$} & $1$ & $2$ & $3$ & $4$ & $5$ & $6$ \\ 
  \hline
(1,1) & 0.19 &  0.06 &  0.22 &  0.18 &  0.23 &  0.12 \\
(1,2) & 0.21 &  0.19 &  0.28 &  0.09 &  0.06 &  0.17\\
(1,3) & 0.19 &  0.13 &  0.18 &  0.19 &  0.28 &  0.03\\
(1,4) & 0.3  &  0.24 &  0.17 &  0.07 &  0.07 &  0.15\\
(1,5) & 0.03 &  0.05 &  0.51 &  0.01 &  0.25 &  0.15\\
(1,6) & 0.22 &  0.14 &  0.13 &  0.16 &  0.21 &  0.14\\
\vdots & \vdots & \vdots & \vdots & \vdots &\vdots & \vdots \\
(6,1) & 0.03 &  0.07 &  0.21 &  0.21 &  0.32 &  0.16\\
(6,2) & 0.18 &  0.13 &  0.35 &  0.1  &  0.16 &  0.08\\
(6,3) & 0.21 &  0.08 &  0.18 &  0.12 &  0.13 &  0.28\\
(6,4) & 0.18 &  0.05 &  0.19 &  0.36 &  0.14 &  0.08\\
(6,5) & 0.31 &  0.14 &  0.3  &  0.07 &  0.16 &  0.02\\
(6,6) & 0.09 &  0.29 &  0.21 &  0.16 &  0.01 &  0.24\\
 \hline
 \end{tabular}
\caption{Best action probabilities $a_t(y_t|x_t,x_{t-1})$ for $\boldsymbol{Q}_x$ in Table \ref{tb:6sQx}, $\beta=[\frac{1}{6}, \dots, \frac{1}{6}]$ and $\lambda=3$.}
    \label{tb:6sAt}
\end{table}


\vspace{-0.5cm}
\subsection{Numerical Results for GeoLife Dataset}
Next, we present the simulation results on the GeoLife dataset \cite{GeoLife,GeoLife2}, which contains 182 user's GPS trajectories collected by Microsoft Research Asia. GeoLife trajectories are recorded densely, e.g., every $1 \sim 5$ seconds or every $5 \sim 10$ meters per point \cite{GeoLife2}. In our experiments, we focus on the high-density areas which represent the important stops for the users. Hence, we use a density-based data mining algorithm, namely DBSCAN (density-based spatial clustering of applications with noise) \cite{DBSCAN} to cluster the raw GPS data into the important stops of the user trajectory. We obtain a 16-cluster representation of the user-016's data, i.e., $\mathcal{W}=16$, by applying DBSCAN algorithm to the 51 trajectories of user-016 provided in GeoLife dataset. For the implementation of our MDP approach in the clustered dataset, center-points of the clusters represent user locations $X_t \in \mathcal{W}$, and the trajectories through the clusters represent user's state transitions. We use Euclidean distance between the true and released user cluster centers as the distortion measure.

Assuming that the user mobility forms a first-order Markov chain, we generate a transition probability matrix $\boldsymbol{\mathcal{Q}}^{016}_x$ from the user-016's trajectories. That is, we assume the user location $X_t$ at time $t$ depends only on the previous location $X_{t-1}$, and we find the empirical probabilities of transitions between locations. After the generation of $\boldsymbol{\mathcal{Q}}^{016}_x$, implementation of PDRP-IDC, PDRP-ADC or the myopic policy is the same as in the synthetic data case. To obtain the optimal policies, we train two fully connected feed-forward DNNs, representing the actor and critic networks, respectively, by using ADAM optimizer. Both networks contain two hidden layers each with 3000 nodes. While all the hidden layers have ReLU activation, the output layers of the actor and critic networks have tanh and Softmax activations, respectively. We obtain the PUT curves by averaging the total information leakage for the corresponding distortion constraint over a time horizon of $n=600$ for 1000 Monte Carlo roll-outs.

Note that the mutual information computed based on the first-order Markov assumption, used by our approach to obtain the PDRP, may not correspond to the real information leakage. Since we do not know the underlying "true" statistics of the data, we examine the effectiveness of the proposed algorithms using an adversary which tries to predict the user's current true location from past released locations in an online manner.
The predictor consists of an LSTM recurrent neural network layer with 200 nodes and a dropout of 0.5, which is followed by a fully connected hidden layer of 200 nodes with ReLu activation, and a fully connected output layer with Softmax activation. We train the predictor on the released distorted locations with the goal of minimizing the categorical cross-entropy between the estimated and true current locations by utilizing ADAM optimizer.

In Table \ref{tab:PredIDC}, we show the adversary's cross-entropy loss for predicting user-016's true locations from their distorted versions released by PDRP-IDC at various PUT points. Here, $m$ is the LSTM based adversary's look-back memory. For both $m=1$ and $m=5$, Table \ref{tab:PredIDC} shows that the cross-entropy loss decreases as the average information leakage increases. In Table \ref{tab:PredIDC}, there is a decrease in the adversarial loss for $m=5$ compared to $m=1$, which means that the first-order Markov assumption may not be valid for the data as the adversary benefits from considering information further in the past. To understand the benefit of releasing distorted data better, we also obtained the cross-entropy loss of the adversary when it predicts the current location by observing the past true locations. When the privacy is not preserved, the adversary's cross-entropy loss is $0.36$ for $m=1$ and $0.28$ for $m=5$, which is much lower than the privacy preserved case as expected.


\tabcolsep=0.22cm
\begin{table}[pt]
\centering
\begin{tabular}{|l|l|l|l|l|l|}
\hline
\multicolumn{3}{|l|}{Instantaneous Distortion Const.:}  & 15 km   & 5 km  & 3 km \\ \hline
\multicolumn{1}{|c|}{\multirow{3}{*}{PDRP-IDC}} & \multicolumn{2}{l|}{Avg. Info. Leakage}   & 0.18  & 0.39  & 0.53 \\ \cline{2-6} 
\multicolumn{1}{|c|}{}                          & \multirow{2}{*}{Cross-entropy Loss} & m=1 & 1.05 & 0.66 & 0.52 \\ \cline{3-6} 
\multicolumn{1}{|c|}{}   &     & m=5 &  0.46    & 0.40     &   0.35   \\ \hline
\end{tabular}
\caption{Cross-entropy loss of the predictor for certain PUT levels of PDRP-IDC.}
\label{tab:PredIDC}
\end{table}



\tabcolsep=0.16cm
\begin{table}[pb]
\centering
\begin{tabular}{|l|l|l|l|l|l|}
\hline
\multicolumn{3}{|l|}{Average Distortion Const.:}                             & 9 km   & 5.7 km & 1.7 km \\ \hline
\multicolumn{1}{|c|}{\multirow{3}{*}{PDRP-ADC}} & \multicolumn{2}{l|}{Avg. Info. Leakage}   & 0.11 & 0.20  & 0.35 \\ \cline{2-6} 
\multicolumn{1}{|c|}{}                          & \multirow{2}{*}{Cross-entropy Loss} & m=1 & 1.30 & 1.25 & 0.90 \\ \cline{3-6} 

\multicolumn{1}{|c|}{}                          &                      & m=5 &  0.78    & 0.73     &   0.67   \\ \hline
\multirow{3}{*}{Myopic PDRP}                    & \multicolumn{2}{l|}{Avg. Info. Leakage}   & 0.27 & 0.33 & 0.50  \\ \cline{2-6} 
                                                & \multirow{2}{*}{Cross-entropy Loss} & m=1 & 1.10 & 0.99 & 0.82 \\ \cline{3-6} 
             &    & m=5 & 0.52  & 0.48  &   0.45   \\ \hline
\end{tabular}
\caption{Cross-entropy loss of the predictor for certain PUT levels of PDRP-ADC and myopic policy.}
\label{tab:PredADC}
\end{table}

In Table \ref{tab:PredADC}, we show the adversary's prediction performance against PDRP-ADC and the myopic policy at various PUT points. For the same average distortion constraints, the adversary has higher cross-entropy loss of predicting true locations when they are distorted by PDRP-ADC rather than the myopic policy for both $m=1$ and $m=5$. Hence, considering the temporal correlations in the trajectory preserves PDRP-ADC's advantage over the myopic policy even when the adversary has a less strict Markov assumption on the true location distribution than both policies.


\begin{figure}[pt]
\centering
\subfloat{}{\label{fig:MapsX}
\includegraphics[width=7.2cm]{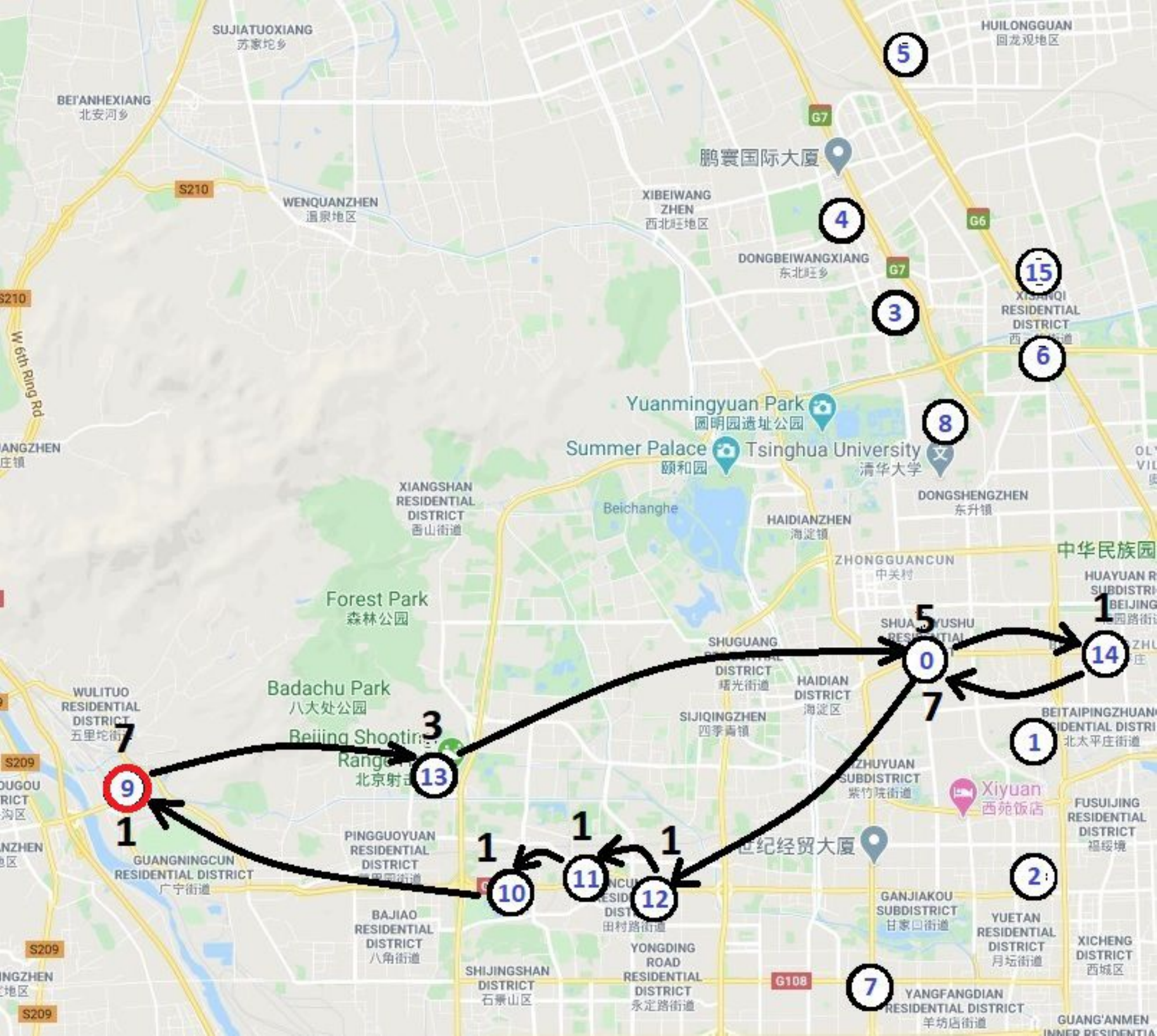}%
\caption*{(a)}
}
\vfill
\subfloat{}{\label{fig:MapsY}
\includegraphics[width=7.2cm]{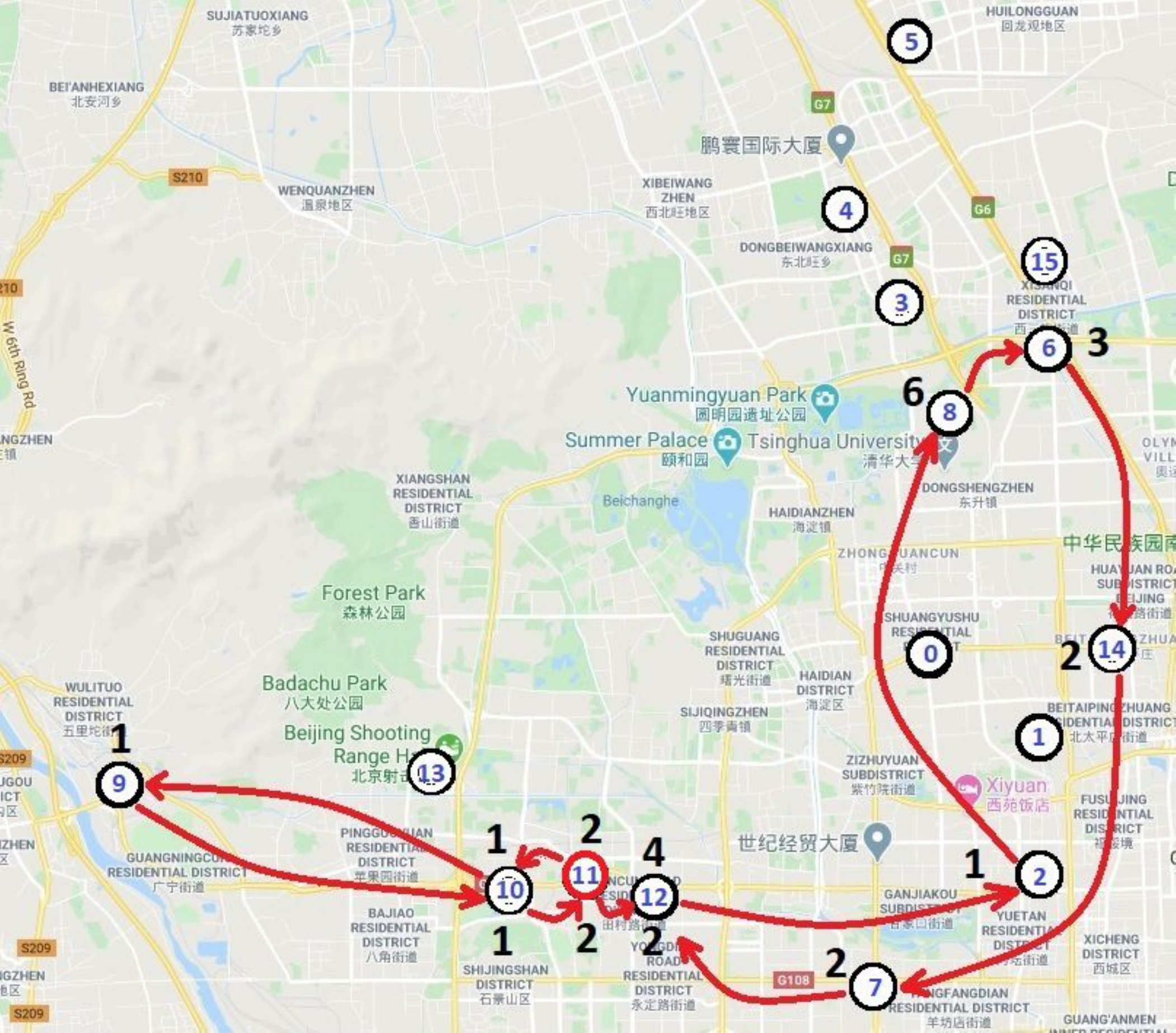}%
\caption*{(b)}
}
\caption{True (a) and the distorted (b) trajectory of user-016 by PDRP-ADC for $\mathcal{W}=16$, $\lambda=1$ and $\hat{D}=5$km.}
\label{fig:MapsTrajectories}
\end{figure}

To understand the true and released location trajectories better, we provide a toy example in which we apply PDRP-ADC to previously clustered user-016 trajectories for $\mathcal{W}=16$, $\lambda=1$ and $\hat{D}=5$km. An example for the true trajectory of the user is shown in Fig. \ref{fig:MapsTrajectories}a, where the numbered circles are the cluster center-points with the corresponding cluster numbers in blue, black numbers represent how many steps the user takes in that cluster, the black arrows show the direction of the movement and the movement starts from the red circled cluster $9$. For instance, Fig. \ref{fig:MapsTrajectories}a represents the true trajectory $\{9,9,9,9,9,9,9,13,13,13,0,0,0,0,0,14,0,0,0, \dots\}$. The distorted version of the trajectory in Fig. \ref{fig:MapsTrajectories}a is depicted in Fig. \ref{fig:MapsTrajectories}b, where the movement starts from the red circled cluster $11$ and the red arrows show the direction of movement. The released trajectory can be deduced from the map in Fig. \ref{fig:MapsTrajectories}b as $\{11,11,10,9,10,11,11,12,12,12,12,2,8,8,8,8,8,8,6,\dots\}$. These figures show that the released locations by PDRP-ADC follow a different path from the true locations for privacy concerns, while the distortion constraint is satisfied.  


\section{Conclusions}
\label{sec:Conc}
We have studied the PUT of time-series data using mutual information as a privacy measure. Having identified some properties of the optimal policy, we proposed information theoretically optimal online PDRPs under instantaneous and average distortion constraints, which represent utility constraints, and solved the PUT problem as an MDP. Due to continuous state and action spaces, it is challenging to characterize or even numerically compute the optimal policy. We overcome this difficulty by employing advantage actor-critic deep RL as a computational tool. Then, we applied the theoretical approach which we introduced for time-series data privacy into the location trace privacy problem. Utilizing DNNs, we numerically evaluated the PUT curve of the proposed PDRPs under both instantaneous and average distortion constraints for both synthetic data and GeoLife GPS trajectory dataset. We compared the results with the myopic location release policy introduced recently in \cite{Ravi}, and observed the effect of considering temporal correlations on information leakage-distortion performance. We also examined the effectiveness of our Markov assumption by testing the proposed policies using an LSTM-based predictor network which represents the adversary with adjustable  memory. According to the simulation results, we have seen that the proposed data release policies provide significant privacy advantage, especially when the user trajectory has higher temporal correlations. Even though higher privacy leakage was observed for larger adversary memory, proposed policies outperformed myopic policy.


\appendices
\section{Proof of Theorem 1}
\label{apx:1}
The proof of Theorem \ref{prop:1} relies on the following lemmas and will be presented later.

\begin{lemma}
For any $\boldsymbol{q} \in \mathcal{Q}_H$, 
\begin{align}
    I^{\boldsymbol{q}}(X^n;Y^n) \geq \sum \limits_{t=1}^{n} I^{\boldsymbol{q}}(X_t,X_{t-1};Y_t|Y^{t-1})
\end{align}
with equality if and only if $\boldsymbol{q} \in \mathcal{Q}_S$.
\label{lem:1}
\end{lemma}

\begin{IEEEproof}
For any $\boldsymbol{q} \in \mathcal{Q}_H$,
    \begin{align}
        I^{\boldsymbol{q}}(X^n;Y^n) &= \sum\limits_{t=1}^{n}I^{\boldsymbol{q}}(X^t;Y_t|Y^{t-1}) \label{eq:10}\\
        &\geq \sum\limits_{t=1}^{n}I^{\boldsymbol{q}}(X_t,X_{t-1};Y_t|Y^{t-1}), \label{eq:11}
    \end{align}
where (\ref{eq:10}) follows from (\ref{eq:2}), and (\ref{eq:11}) from the fact that mutual information cannot be negative. 
\end{IEEEproof}


\begin{lemma}
For any $\boldsymbol{q}_h \in \mathcal{Q}_H$, there exists a policy $\boldsymbol{q}_s \in \mathcal{Q}_S$ such that
\begin{align}
    \sum \limits_{t=1}^{n} I^{\boldsymbol{q}_h}(X_t,X_{t-1};Y_t|Y^{t-1})
  = \sum \limits_{t=1}^{n} I^{\boldsymbol{q}_s}(X_t,X_{t-1};Y_t|Y^{t-1}), \label{eq:lem2_IC}
\end{align}
for both cases where $\boldsymbol{q}_h$ and $\boldsymbol{q}_s$ satisfy an instantaneous distortion constraint $d(X_t,Y_t)\leq\bar{D}$, and average distortion constraints $\mathbb{E}^{\boldsymbol{q}_h}\Big[\frac{1}{n}\sum\limits_{t=1}^n\Big] \leq \bar{D}$ and $\mathbb{E}^{\boldsymbol{q}_s}\Big[\frac{1}{n}\sum\limits_{t=1}^n\Big] \leq \bar{D}$, respectively. 
\label{lem:2}
\end{lemma}

\begin{IEEEproof}
For any $\boldsymbol{q}_h \in \mathcal{Q}_H$,  we choose the policy $\boldsymbol{q}_s \in \mathcal{Q}_S$ such that 
    \begin{align}
         \hspace{-0.3cm} q_t^s(y_t|x_t,x_{t-1},y^{t-1}) \hspace{-0.1cm} =  \hspace{-0.1cm} P^{\boldsymbol{q}_h}_{Y_t|X_t,X_{t-1},Y^{t-1}}(y_t|x_t,x_{t-1},y^{t-1}),  \hspace{-0.2cm} \label{eq:13}
    \end{align}
and we show that $P^{\boldsymbol{q}_h}_{X_t,X_{t-1},Y^{t}}=P^{\boldsymbol{q}_s}_{X_t,X_{t-1},Y^{t}}$. Then, $I^{\boldsymbol{q}_h}(X_t,X_{t-1};Y_t|Y^{t-1})=I^{\boldsymbol{q}_s}(X_t,X_{t-1};Y_t|Y^{t-1})$ holds, which proves the statement in Lemma \ref{lem:2}.
The proof of the equality $P^{\boldsymbol{q}_h}_{X_t,X_{t-1},Y^{t}}=P^{\boldsymbol{q}_s}_{X_t,X_{t-1},Y^{t}}$ requires the proof of $P^{\boldsymbol{q}_h}_{X_t,X_{t-1},Y^{t-1}}=P^{\boldsymbol{q}_s}_{X_t,X_{t-1},Y^{t-1}}$ which is derived by induction as follows,
    \begin{align}
         &P^{\boldsymbol{q}_h}(x_{t+1},x_t,y^t) \nonumber \\
        & = \hspace{-0.3cm} \sum_{ \substack{ x_{t-1} \in \mathcal{W}}} \hspace{-0.3cm}  q_x(x_{t+1}|x_t)q_t^h(y_t|x_t,x_{t-1},y^{t-1}) P^{\boldsymbol{q}_h}(x_{t},x_{t-1},y^{t-1}) \nonumber \\
        & = \hspace{-0.3cm} \sum_{ \substack{ x_{t-1} \in \mathcal{W} }} \hspace{-0.3cm}  q_x(x_{t+1}|x_t)q_t^s(y_t|x_t,x_{t-1},y^{t-1})P^{\boldsymbol{q}_s}(x_{t},x_{t-1},y^{t-1}) \nonumber\\
        & = P^{\boldsymbol{q}_s}(x_{t+1},x_t,y^t),
    \end{align}
where (\ref{eq:13}) holds, and $P^{\boldsymbol{q}_h}_{X_1}(x)=p_{x_1}(x)=P^{\boldsymbol{q}_s}_{X_1}(x)$ is used for the initialization of the induction. 

Having shown that the equality $P^{\boldsymbol{q}_h}_{X_t,X_{t-1},Y^{t-1}}=P^{\boldsymbol{q}_s}_{X_t,X_{t-1},Y^{t-1}}$ and (\ref{eq:13}) hold, the proof of $P^{\boldsymbol{q}_h}_{X_t,X_{t-1},Y^{t}}=P^{\boldsymbol{q}_s}_{X_t,X_{t-1},Y^{t}}$ is straightforward:
\begin{align}
        P^{\boldsymbol{q}_h}(x_{t},x_{t-1},y^t)  &= q_t^h(y_t|x_t,x_{t-1},y^{t-1})P^{\boldsymbol{q}_h}(x_{t},x_{t-1},y^{t-1}) \nonumber\\
        & = q_t^s(y_t|x_t,x_{t-1},y^{t-1})P^{\boldsymbol{q}_s}(x_{t},x_{t-1},y^{t-1}) \nonumber\\
        & = P^{\boldsymbol{q}_s}(x_{t},x_{t-1},y^t). \label{eq:prf}
\end{align}
Following (\ref{eq:prf}), the equality $I^{\boldsymbol{q}_h}(X_t,X_{t-1};Y_t|Y^{t-1})=I^{\boldsymbol{q}_s}(X_t,X_{t-1};Y_t|Y^{t-1})$ holds, and the integration of the instantaneous distortion constraint into the additive mutual information is straightforward and does not affect the optimality, and hence, (\ref{eq:lem2_IC}) holds.

Furthermore, we show that there is no loss of optimality in including the average distortion constraint into the mutual information optimization when the policy is chosen according to (\ref{eq:13}), as follows:
\begin{align}
    \mathbb{E}^{\boldsymbol{q}_h}[d(X_t,Y_t)]
    &=  \hspace{-0.3cm} \sum \limits_{\substack{y^t \in \mathcal{W}^t, \\ x_t,x_{t-1} \in \mathcal{W}}}P^{\boldsymbol{q}_h}(x_t,x_{t-1},y^t)d(x_t,y_t) \label{eq:ADCproof1}\\
    &=  \hspace{-0.3cm} \sum \limits_{\substack{y^t \in \mathcal{W}^t, \\ x_t,x_{t-1} \in \mathcal{W}}}P^{\boldsymbol{q}_s}(x_t,x_{t-1},y^t)d(x_t,y_t), \label{eq:ADCproof2}\\
    &=\mathbb{E}^{\boldsymbol{q}_s}[d(X_t,Y_t)]
\end{align}
where (\ref{eq:ADCproof1}) follows from the history independence of $d(X_t,Y_t)$, and (\ref{eq:ADCproof2}) from (\ref{eq:prf}). Following the linearity of expectation, the average distortion constraint can be written in an additive form, and hence, (\ref{eq:lem2_IC}) holds.

\end{IEEEproof}
\vspace{0.5cm}
\begin{IEEEproof}[Proof of Theorem \ref{prop:1}]
Following Lemmas \ref{lem:1} and \ref{lem:2}, for any $\boldsymbol{q}_h \in \mathcal{Q}_H$, there exists a $\boldsymbol{q}_s \in \mathcal{Q}_S$ such that \begin{align}
    I^{\boldsymbol{q}_h}(X^n;Y^n) \geq I^{\boldsymbol{q}_s}(X^n;Y^n).
\end{align}

Hence, there is no loss of optimality in using the time-series data release policies of the form $q_t^s(y_t,|x_t,x_{t-1},y^{t-1})$, and information leakage and the average distortion constraint reduce to (\ref{eq:theo1MI}) and (\ref{eq:theo1Dist}), respectively.
\end{IEEEproof}

\ifCLASSOPTIONcaptionsoff
  \newpage
\fi



\bibliographystyle{IEEEtran}
\bibliography{IEEEexample}

\end{document}